\documentclass[a4paper, accepted=2024-09-17]{quantumarticle}

\pdfoutput=1

\usepackage{caption}
\usepackage{subcaption}
\usepackage[section]{placeins}
\usepackage[numbers]{natbib}

\usepackage{macros}
\graphicspath{{figs/}}

\newcommand{\mypar}[1]{\textbf{#1\ }}

\newcommand{\tildenormal}{\raisebox{0.5ex}{\texttildelow}}

\newcommand{\subsub}[1]{\subsubsection{#1}}

\makeatletter
\renewcommand{\fnum@figure}{Figure \thefigure}
\makeatother

\begin{document}

\title{Tight and Efficient Gradient Bounds for Parameterized Quantum Circuits}
\author{Alistair Letcher}
\author{Stefan Woerner}
\author{Christa Zoufal}
\affiliation{IBM Quantum, IBM Research Europe -- Zurich}

\begin{abstract}
The training of a parameterized model largely depends on the landscape of the underlying loss function.
In particular, vanishing gradients (also known as \textit{barren plateaus}) are a central bottleneck in the scalability of variational quantum algorithms (VQAs),
and are known to arise in various ways, from circuit depth and hardware noise to global observables.
However, a caveat of most existing gradient bound results is the requirement of $t$-design circuit assumptions that are typically not satisfied in practice.
In this work, we loosen these assumptions altogether and derive tight upper and lower bounds on loss and gradient concentration for a large class of parameterized quantum circuits and arbitrary observables, which are significantly stronger than prior work. Moreover, we show that these bounds, as well as the variance of the loss itself, can be estimated efficiently and classically
-- providing practical tools to study the loss landscapes of VQA models, including verifying whether or not a circuit/observable induces barren plateaus.
In particular, our results can readily be leveraged to rule out barren plateaus for a realistic class of ansätze and mixed observables, namely, observables containing a non-vanishing local term.
This insight has direct implications for hybrid Quantum Generative Adversarial Networks (qGANs), a generative model that can be reformulated as a VQA with an observable composed of local and global terms. We prove that designing the discriminator appropriately leads to 1-local weights that stay constant in the number of qubits, regardless of discriminator depth. Combined with our first contribution, this implies that qGANs with appropriately chosen generators do not suffer from barren plateaus even at scale -- making them a promising candidate for applications in generative quantum machine learning. We demonstrate this result by training a qGAN to learn a 2D mixture of Gaussian distributions with up to 16 qubits, and provide numerical evidence that global contributions to the gradient, while initially exponentially small, may kick in substantially over the course of training.
\end{abstract}

\maketitle

\section{Introduction}
Gradients that vanish exponentially in the number of system qubits -- also called \emph{barren plateaus} -- have been shown to provide a substantial bottleneck in the training of variational quantum algorithms~\cite{Clean_2018_BarrenPlateaus, Cerezo_2021_costfunct, Cerezo_2021higher_order_bps, holmes2021AnsatzExpressBarrenPlateaus, Wiebe2020Barren, Wang_21_noiseinducedBPs, napp2022quantifying, Uvarov_2021BPs}.
They pose a central obstacle for the scaling of these algorithms to practically relevant problem sizes, and have been shown to arise for a variety of reasons: ansatz depth or expressivity~\cite{Clean_2018_BarrenPlateaus, holmes2021AnsatzExpressBarrenPlateaus}, entanglement~\cite{Wiebe2020Barren}, unital hardware noise~\cite{Wang_21_noiseinducedBPs}, and loss functions induced by global observables, namely, observables acting on most system qubits~\cite{Cerezo_2021_costfunct}.
Despite positive guarantees in a number of settings \cite{pesah2020absenceQCNNs, sharma2020trainabilityDissipative, Grant_2019BPInitialization, rudolph2022synergy, Zhao-2021-zx, wang2023uniform, zhang2022gaussian, BenedettiF-Divergences21}, these results naturally beckon a finer understanding of the conditions under which VQAs are scalable to intermediate- and large-scale problems ranging across optimization \cite{FarhiQAOA14, abbas2023quantum}, machine learning \cite{19HavSupervisedLearning}, and chemistry \cite{PeruzzovarEigenSolverPhotonic14}.

While ansatz and backend may be chosen with some degree of freedom, the observable is often intrinsically connected to the model at hand. 
Notably, a large spectrum of VQA applications
involve 
\textit{mixed} 
observables which decompose as a sum of \textit{local} and \textit{global} terms. 
As proven in \cite{Cerezo_2021_costfunct}, the variance of local terms is typically at most polynomially small, whereas global terms often induce an exponentially vanishing variance. Furthermore, \cite{Uvarov_2021BPs} showed that local and global terms contribute independently to the gradient. Both results make use of local 2-design assumptions that are not often satisfied in practice, and it remains unclear whether mixed observables induce barren plateaus in realistic settings. 
Indeed, it is known that exact and approximate 2-designs on $m$ qubits require $\mathcal{O}\left(m^2\right)$ and $\mathcal{O}\left(m\right)$ gates to be implemented respectively \cite{Dankert09exactandapprox_2designs} -- but these are asymptotic statements and do not guarantee a fixed unitary to behave as an (approximate) 2-design. A numerical study of the proximity of local blocks to (approximate) 2-designs can be found in \cite{Uvarov_2021BPs}.
Furthermore, local restrictions or approximations of the underlying loss function have widely been suggested as potential remedy \cite{BenedettiF-Divergences21, rudolph2023trainability, sharma2020trainabilityDissipative, bravoprieto2020variational, Volkoff_2021_LargeGradients, Khatri2019quantumassisted, HolmesVariationalFastForwarding20}. However, while local loss functions can help avoid vanishing gradients, they are also likely to introduce undesirable local minima in the loss landscape \cite{AnschuetzQuantumTraps22, CerezoVariationalQSE22} and fail to capture all system correlations.

Our work is centered around two main contributions.
First, we derive tight loss and gradient bounds for a large class of parameterized circuits and arbitrary observables (Theorem \ref{th:bounds}), \textit{without} any $t$-design assumptions. These bounds, as well as the variance of the loss itself, can be estimated efficiently and classically, and are significantly stronger than prior work (Figure~\ref{fig:light-cones-bounds}).
This involves proof techniques and intuitions that are more practical and elementary, and can readily be leveraged for other classes of ansätze. In particular, our results imply that mixed observables do not induce barren plateaus (Figure~\ref{fig:locality_variance}) -- and that approximating a true loss function by its local counterpart can only \textit{increase} the concentration of gradients, besides leading to potentially undesirable local minima. It is nonetheless important to emphasize that the absence of barren plateaus is a \textit{necessary} condition to train parameterized circuits -- but in no way \textit{guarantees} that an optimal solution will be found, as is generically true in non-convex optimization.

Second, we provide a bridge between our gradient bounds and the field of generative modeling by proving that for hybrid qGANs \cite{Zoufal2019qGANs, SITU2020193qGANs, killoran2018qgans, romero21qGAN, Zeng_2019LearningInferenceonqGANs}, the contribution of 1-local terms in the overall loss function stays \emph{constant} in the number of qubits, even for classical discriminators of \emph{arbitrary depth} (Theorem \ref{th:qgan}). Together, our results imply that qGANs with shallow generators do not suffer from barren plateaus. Since most other GQML algorithms do suffer from exponentially vanishing gradients, our results suggest that qGANs are promising candidates for scalable GQML algorithms. We empirically demonstrate this by successfully training a qGAN to learn 2D mixtures of Gaussians with up to 16 qubits, with results which are on par with classical GANs \cite{letcher2019stable} -- despite a significantly smaller number of parameters.

The remainder of this paper is organized as follows.
We introduce relevant concepts and related work in Section \ref{sec:background}.
We then present our main contributions in Section \ref{sec:results}. The first sub-section \ref{sec:general-theory} introduces Theorem \ref{th:bounds}, which provides tight gradient bounds for arbitrary observables and a large class of parameterized quantum circuits. Several extensions and practical implications thereof are discussed.
Next, we leverage these results in sub-section \ref{sec:qgans} to prove that the gradients of hybrid qGANs do not vanish exponentially for arbitrarily deep discriminators and logarithmically shallow generators. Following this, Section \ref{sec:qgans-experiments} presents numerical experiments that support the previously discussed theoretical insights.
Finally, Section \ref{sec:conclusion} gives a conclusion and an outlook for future research.

\section{Background}\label{sec:background}

\subsection{Variational Quantum Algorithms} In this paper, we study optimization problems that can be formulated as a VQA \cite{cerezo-vqas-2021}, i.e., that can be reduced to the minimization of a loss function
\begin{equation}\label{eq:vqa}
\cL(\T) = \Tr(U(\T) \rho U^\dagger(\T)  H),
\end{equation}
induced by an $n$-qubit parameterized quantum circuit $U(\T)$, an initial state $\rho$, and a Hermitian observable $H$.

The set of problems that can be formulated as VQAs find a variety of applications. Prime examples include chemical ground state problems
\cite{KandalaHWEffVQE17, PeruzzovarEigenSolverPhotonic14}, black box and polynomial binary optimization
\cite{Zoufal2023VariationalQuantumBlackBox, MatsumoriQUBOBlackBox22, IzawaContinuousBlackBox22, alcazar2022geo} (Appendix \ref{app:equivalence}), and distribution learning algorithms such as qGANs \cite{Zoufal2019qGANs, huqGANs2019, lloyd2qGANs18, romero21qGAN, killoran2018qgans}, which we develop in Section~\ref{sec:qgans}.

It is worth noting that these problems are often governed by \textbf{mixed} Hermitian observables, as defined in the following column, rather than purely \textbf{local} interactions. For example in quantum chemistry, this includes hydrogen chains \cite{SokolovQOrbitalOptUCC20, PerezLearningMeasure21, HachmannMultiRefCorrMolecules06, LimacherMeanField13, MottaManyElectronProbRealMat17}, vibrational bosonic systems \cite{OllitraultHardwareEffQAVirationalStructure20, McArdleDigitalQuantumSimVibrations19, SawayaVibrationalSpectroscopy21}, and downfolded electronic Hamiltonians \cite{KowalskiDimensionalityReduction21, HuangDownfoldedHamiltonians23, bauman2023coupled}.
Even many-body systems \cite{IsingFormulationsLucas_2014, Barahona_1982Isingspinglass, StanisicFermi_hubbard22, CadeFermi-Hubbard20, JiangMany-Body18} can exhibit a mixed nature when considered in the first quantization formulation \cite{ollitrault2022quantumGridVTE} or after application of certain fermion-to-qubit mappings \cite{JordanWigner1928, ParityMappingBravyi-Kitaev12, FermionicQuantumComputationBRAVYI20022}, which are required to map fermionic creation and annihilation operators to Pauli operators.

\subsection{Barren Plateaus} A loss function $\cL(\T)$ is defined to exhibit a barren plateau if, for all $\T_k$, $\Var_\T\left[ \partial_k \cL \right] \in O(1/b^n)$ for some $b > 1$, where the distribution on $\T$ is uniform. \cite{Arrasmith_2022} have proven that, in the VQA setting, this is equivalent to an exponential concentration of the \textit{loss function itself}, namely, $\Var\left[ \cL \right] \in O(1/b^n)$ for some $b > 1$. 
This simplifies the analysis of barren plateaus, although our main result (Theorem \ref{th:bounds}) provides bounds on the concentration of both loss and gradients for completeness.

\begin{figure*}[t!]
\centering
\begin{subfigure}[c]{0.4\linewidth}
\subcaption{}
\includegraphics[width=\linewidth]{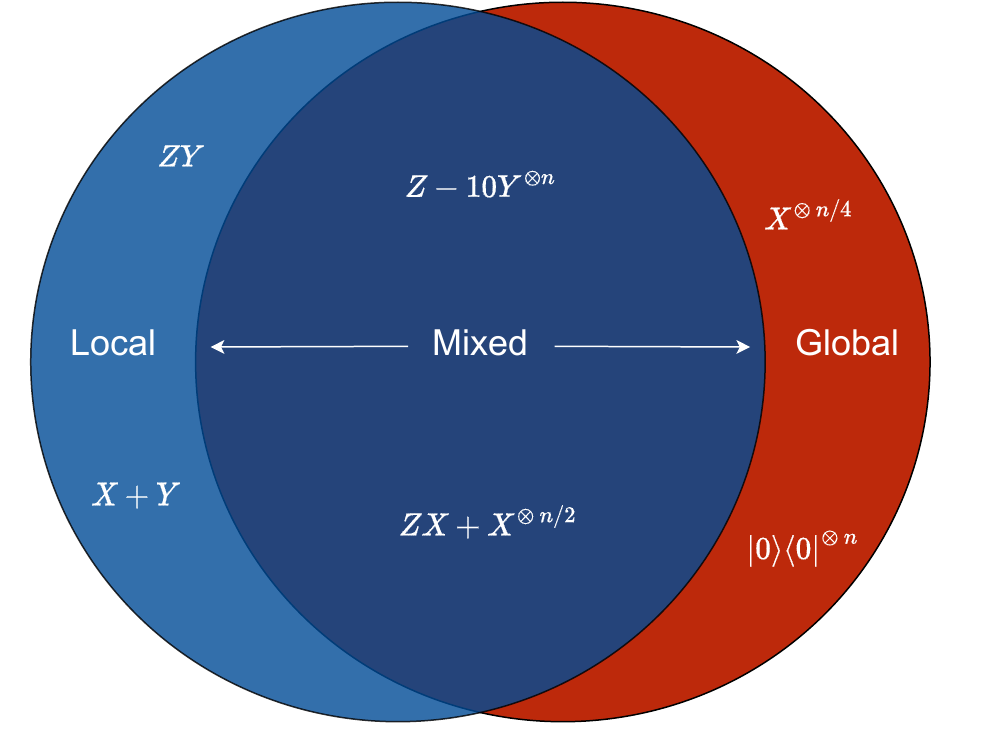}
\end{subfigure} \hspace*{30pt}
\begin{subfigure}[c]{0.4\linewidth}
\subcaption{}
\includegraphics[width=\linewidth]{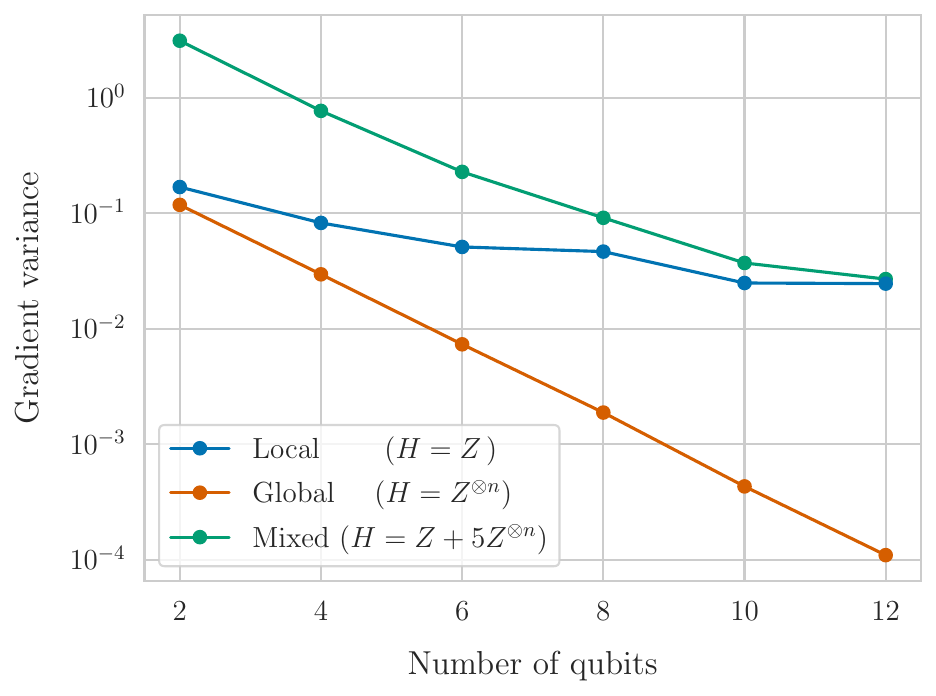}
\end{subfigure}
\caption{(a) Illustration and examples of local vs.~mixed vs.~global observables. All observables act on $n$ qubits, but we omit the identity terms for convenience. (b) Variance of the first-parameter gradient $\Var \left[ \partial_1 \cL \right]$ for a local, global, and mixed observable paired with an EfficientSU2 circuit of logarithmic depth. The mixed observable has a larger gradient variance than either of its Pauli terms (Theorem~\ref{th:bounds}), and does not induce barren plateaus (Corollary~\ref{cor:su2}).}
\label{fig:locality_variance}
\end{figure*}

\subsection{Locality} To describe whether a given observable induces barren plateaus, various notions of \textit{locality} have been introduced \cite{Cerezo_2021_costfunct, Uvarov_2021BPs, rudolph2023trainability}.
Our central contribution, Theorem \ref{th:bounds}, provides bounds for \textit{any observable}, and is thus independent from locality.
However, in order to exclude barren plateaus for specific circuit-observable pairs, as in Corollary \ref{cor:su2}, a notion of locality is required.
In this work, we say that $H$ is \textbf{local} if it acts non-trivially on a constant number $O(1)$ of qubits, and \textbf{global} otherwise. This is referred to as \textit{algebraic locality}, which is independent of circuit topology -- though Corollary \ref{cor:su2} also extends to other notions of \textit{topological locality} and \textit{low-bodiness} \cite{rudolph2023trainability}.

This notion of locality can be refined by decomposing $H$ into a sum of Pauli strings as $H = \sum_\al c_\al P_\al$, with $c_\al \in \R$ and $\al \in \{0, 1, 2, 3\}^n$, where a \textit{Pauli string} $P_\al$ is defined by $P_\al = \bigotimes_{i=1}^n \sig_{\al_i}$ and $\sig_{\al_i} \in \{I, X, Y, Z\}$ are the Pauli matrices. We can then define $P_\al$ to be \textit{$k$-local} if it contains exactly $k$ non-identity matrices, or formally, if $\abs{\al} = k$ in the $L_0$ pseudo-norm ($\abs{\al} \coloneqq \abs{ \{ \al_i \neq 0 \} }$). We can then define an observable to be \textbf{mixed} if it contains \textit{at least one} local term $P_\al$ with a non-vanishing coefficient $c_\al \in \Omega(1/\poly n)$. Note that we do not require observables to have a polynomial number of terms or a polynomially bounded spectral norm, as often assumed in the literature. These distinctions are illustrated, with examples, in Figure \ref{fig:locality_variance}\textcolor{red!80!black}{a}.

\subsection{Related Work} Most existing results focus on barren plateaus in circuits that satisfy certain $t$-design assumptions, either exactly \cite{Clean_2018_BarrenPlateaus, Cerezo_2021_costfunct, Wiebe2020Barren, napp2022quantifying, Uvarov_2021BPs, leone2022practical} or approximately \cite{holmes2021AnsatzExpressBarrenPlateaus, larocca2022diagnosing, fontana2023adjoint, ragone2023unified}.
In particular, \cite[Theorem 1]{Cerezo_2021_costfunct} have proven the existence of barren plateaus whenever $H$ is a specific type of global observable, such as a projector, and $U$ is an Alternating Layered Ansatz forming a local 2-design. Conversely, the gradient is at most polynomially vanishing in $n$ if $H$ is a (type of) spatially local observable, and if $U$ moreover has logarithmic depth in $n$. These results are seminal in the study of observable-induced barren plateaus, but local 2-design assumption are not necessarily satisfied in practice.
As this work will show, relaxing these assumptions leads to more practical bounds with slightly different implications.

By proving that Pauli strings make independent contributions to the gradient, \cite{Uvarov_2021BPs} have extended the \emph{lower} bounds provided by \cite{Cerezo_2021_costfunct} to generic observables -- but also require circuits to form local 2-designs, and do not provide any upper bounds.

\cite{napp2022quantifying} has derived gradient bounds for spatially and algebraically local observables. However, they consider a class of circuits whose entangling gates are chosen randomly according to any measure that forms a 2-design.

For Periodic Structure ansätze, \cite{larocca2022diagnosing} have conjectured that the scaling of the gradient variance is inversely proportional to the dimension of a dynamical Lie algebra associated to the input state. This has recently been proven independently by \cite{fontana2023adjoint} and \cite{ragone2023unified}, 
for observables that are in the Lie algebra associated with the ansatz. While both works provide interesting and novel perspectives on the origins of barren plateaus, they differ from our contributions in requiring circuits to form approximate 2-designs.

\cite{leone2022practical} have analyzed the practical usefulness of the Hardware Efficient Ansatz (HEA), and identify a ``Goldilocks scenario where shallow HEAs could achieve a quantum speedup: QML tasks with data satisfying an area law of entanglement''. This is encouraging, and we hope our contributions can help refine their insights by loosening their $t$-design assumptions.

\cite{Zhao-2021-zx} have ruled out barren plateaus for quantum convolutional networks and tree tensor networks, but these classes of circuits do not apply to problems that are naturally induced by global observables, as they require tracing out almost all qubits before measurement. They also show that a class of hardware-efficient ansätze induces barren plateaus for sufficient depth -- but do not provide lower bounds in order to rule them out for shallow circuits.

Finally, \cite{wang2023uniform} and \cite{zhang2022gaussian} have proposed initialization schemes that rule out barren plateaus without $t$-design assumptions. However, the results of \cite{wang2023uniform} and the main theorem of \cite{zhang2022gaussian} only apply to local observables, while Theorem 4.2 of \cite{zhang2022gaussian} applies to global observables, but only provides non-trivial bounds when the gradient at the origin is sufficiently large. Our work instead focuses on providing tight bounds for generic uniform initialization, a larger class of circuits, and \textit{all observables}.

\begin{figure*}[ht]
\centering
\includegraphics[width=\linewidth]{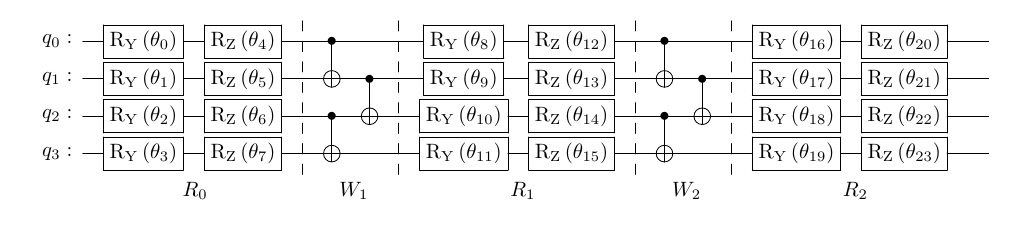}
\caption{EfficientSU2 circuit \cite{efficientsu2} with pairwise entanglement and $(R_Y, R_Z)$ rotation layers, illustrating the circuit structure introduced in Theorem \ref{th:bounds}. After the first layer $R_0$ of single-qubit orthogonal rotations, any Clifford gates $W_k$ and any multi-qubit rotations $R_{P_{kl}}(\T_{kl})$ are allowed.}
\label{fig:qc_thm_1}
\end{figure*}

\section{Results}\label{sec:results}

The central contribution of this section is Theorem~\ref{th:bounds}, where we provide tight upper and lower bounds for loss and gradient concentration, for \textit{arbitrary} observables, and \textit{without any t-design assumptions}. The result is stated for product mixed states to avoid overloading notation and highlight its practical interpretability, but we provide a proof of the generalized result for arbitrary mixed states in Appendix \ref{app:bounds_proof}.

\subsection{Main Theorem}\label{sec:general-theory}

We introduce the circuit class in Definition \ref{def:circuit} below, along with notions of light-cones and orthogonality required to state our main result in Definitions \ref{def:light_cones} and \ref{def:orthogonality}.

\begin{definition}[Circuit class]\label{def:circuit}
Let $\cU$ be the family of parameterized circuits $U(\T)$ of the form
\[ U(\T) = \prod_{k=0}^{K} W_k R_k(\T) \,, \]
where $W_k$ are Clifford gates; $R_k(\T) = \prod_l R_{P_{kl}}(\T_{kl})$ are products of rotation gates generated by arbitrary Pauli strings $P_{kl}$, acting on \textit{any number of qubits}, such that $R_0$  begins with two layers of orthogonal single-qubit rotations (see Figure \ref{fig:qc_thm_1}); and parameters $\T_{kl}$ are independent and initialized uniformly over $[-\pi, \pi]$.
\end{definition}

Moreover, for any circuit $U(\T) \in \cU$, any initial state $\rho$ and any observable $H$, we write
\[ \cL(\T) = \Tr(U(\T) \rho U^\dagger(\T) H) \]
for the induced loss, and similarly, $\cL_\al$ for the loss induced by any Pauli observable  $P_\al$.\newline

Importantly, none of our assumptions restrict the expressivity of the circuit class. In the case of initial orthogonal layers, any parameterized circuit $U(\T)$ can be written as a circuit $U'(\theta, \psi)$ with two initial orthogonal layers parameterized by $\psi$, by setting $\psi = 0$. The same applies to parameters being independent, which can evolve to be (almost) equal over the course of training. Similarly, non-Clifford gates can be approximated using a sequence of parameterized rotation gates with appropriate parameters; specifically, adding $T \propto R_Z(\pi/4)$ to the Clifford group provides a set of universal quantum gates. 

Moreover, these assumptions are necessary to avoid edge cases inducing vanishing gradients. For example, a RealAmplitudes circuit \cite{realamp}, which does not have two initial orthogonal layers, paired with a 1-local $Y$-observable, may have \textit{uniformly zero gradients}. One could loosen the assumption by imposing further restrictions on the \textit{observable} or the \textit{input state}, but these are typically imposed by the problem at hand -- it is thus preferable to impose the weakest possible assumptions on the \textit{ansatz}. We discuss this in full detail in Appendix \ref{app:justification}.

\begin{definition}[Light-cones]\label{def:light_cones} Provided any $m$-parameter circuit $U(\T)$ from Definition \ref{def:circuit}, and any $\T \in [-\pi, \pi]^m$, the \textit{light-cone} $\conealphatheta$ is defined as the number of qubits on which $U^\dagger(\T)P_\al U(\T)$ acts non-trivially.
\end{definition}

By loosening $t$-design assumptions, our results (below) show that gradient concentration is not determined by locality, circuit depth or entanglement \textit{per se}, but by the \textit{interaction} between the observable and circuit, which is captured by the size of light-cones. This has the \textit{a priori} surprising consequence that even a \textit{global} observable does not always induce barren plateaus, as further detailed below. For a more intuitive and visual illustration of how these light-cones arise, and how these notions improve on the bounds in prior work, please refer to Appendix \ref{app:efficient_evaluation}.

\begin{definition}[Orthogonality]\label{def:orthogonality}
If $\rho = \otimes \rho_i$ is a product mixed state, we define a measure of orthogonality
\[ \Om(\rho) = \sum_{\substack{\al \in \{0, 1, 2, 3\}^n \\ \forall i \, : \, \al_i \neq 0, \nu_i}} \Tr(P_\al \rho)^2 = \prod_{i} \, \sum_{\la_i \neq 0,\nu_i} \Tr(\sig_{\la_i} \rho_i)^2 \]
which quantifies the `portion' of $\rho$ which is orthogonal to the first layer of rotations $\otimes_i R_{\sig_{\nu_i}}$.
\end{definition}

As shown in Appendix \ref{app:justification}, an initial state $\rho$ that is aligned with an eigenvector of the first-layer rotations may produce uniformly zero gradients -- motivating us to capture the notion of orthogonality $\Om(\rho)$ provided above. In the special case where $\rho = \ketbra{0}{0}^{\otimes n}$ is the zero initial state, and the first layer of the circuit does not include $Z$-rotations, the initial state is fully orthogonal to the first layer, and we find $\Om(\rho) = 1$.

In Part \textbf{(G)} of Appendix \ref{app:bounds_proof}, we generalize the theorem given below to all mixed states by defining the generalized $\al$-orthogonality $\Om(\rho, \al)$ to be a sum over weights $\Tr(P_\la \rho)^2$ for which $P_\la$ is orthogonal to the first layer, on each qubit line where $P_\al$ does not commute with the second layer.
However, unlike product states, this measure is no longer $\al$-independent, resulting in bounds that are more general but less intuitive and practical. They can, nonetheless, still be estimated efficiently using Monte Carlo sampling if the coefficients $\Tr(P_\la \rho)$ are known.

\begin{theorem}\label{th:bounds}
For any $m$-parameter circuit $U(\T) \in \cU$, any observable $H = \sum_\al c_\al P_\al$, and any mixed state $\rho$, each Pauli term makes an independent contribution to the loss and gradient variances:
\begin{align*}
\Var_\T \big[ \cL \big] &= \sum_\al c_\al^2 \Var_\T \big[ \cL_\al \big] \,, \\
\Var_\T \big[ \nabla \cL \big] &= \sum_\al c_\al^2 \Var_\T \big[ \nabla \cL_\al \big] \,.
\end{align*}
Moreover, writing $\cD$ for the discrete uniform distribution over $\{0, \pi/2\}^m$, each non-trivial term ($\al \neq 0$) can be reduced to a \textbf{discrete} expectation, namely,
\[ \norm{\, \Var_\T \big[ \nabla \cL_\al \big] \,}_{\infty} = \Var_\T \big[ \cL_\al \big] = \E_\cD \big[ \cL^2_\al \big] \,. \]
Finally, for product mixed states $\rho$ (generalized to all states in Appendix \ref{app:bounds_proof}), the loss variance is bounded by
\begin{gather*}
\Om(\rho) \, \E_{\cD} \left[ \  \left( \frac{1}{4} \right)^{\conealphatheta} \ \right] \leq \Var_\T \big[ \cL_\al \big] \leq \E_{\cD} \left[ \ \left( \frac{1}{2} \right)^{\conealphatheta} \ \right] \,,
\end{gather*}
and there are circuit-observable pairs that saturate these inequalities. In particular, combined with the previous equation, at least one partial derivative $\partial_k \cL_\al$ satisfies the lower bound, and all of them satisfy the upper bound.
\end{theorem}

\begin{figure*}[t]
\centering
\begin{subfigure}[t]{0.49\linewidth}
\hspace{30pt}
\subcaption{EfficientSU2}
\hspace{-30pt}
\includegraphics[width=\linewidth]{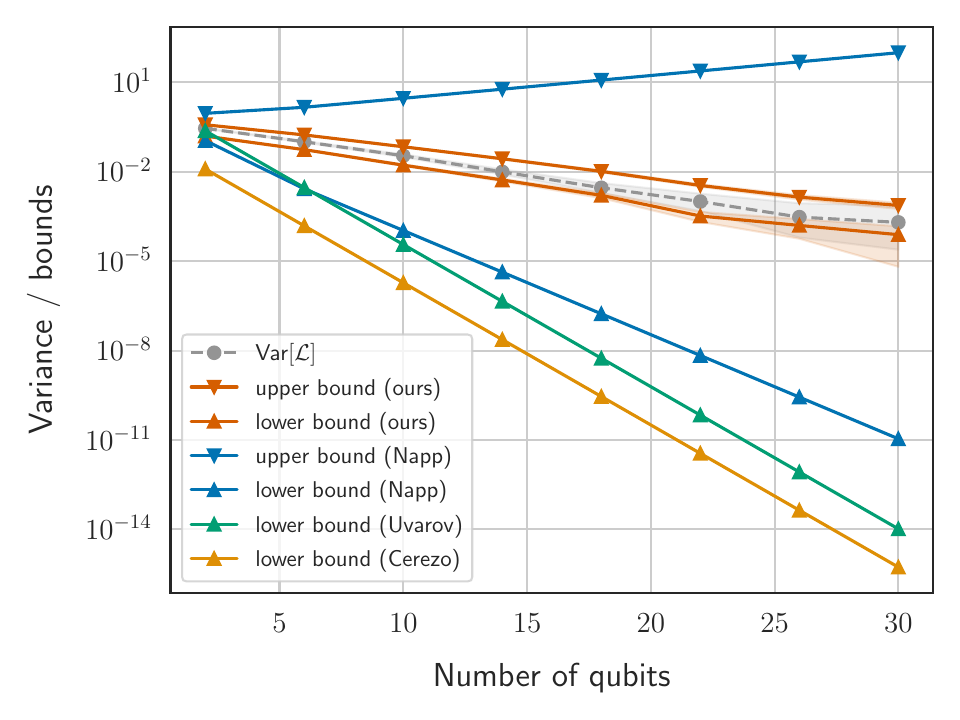}
\end{subfigure} \hfill
\begin{subfigure}[t]{0.49\linewidth}
\hspace{30pt}
\subcaption{Approximate local 2-design}
\hspace{-30pt}
\includegraphics[width=\linewidth]{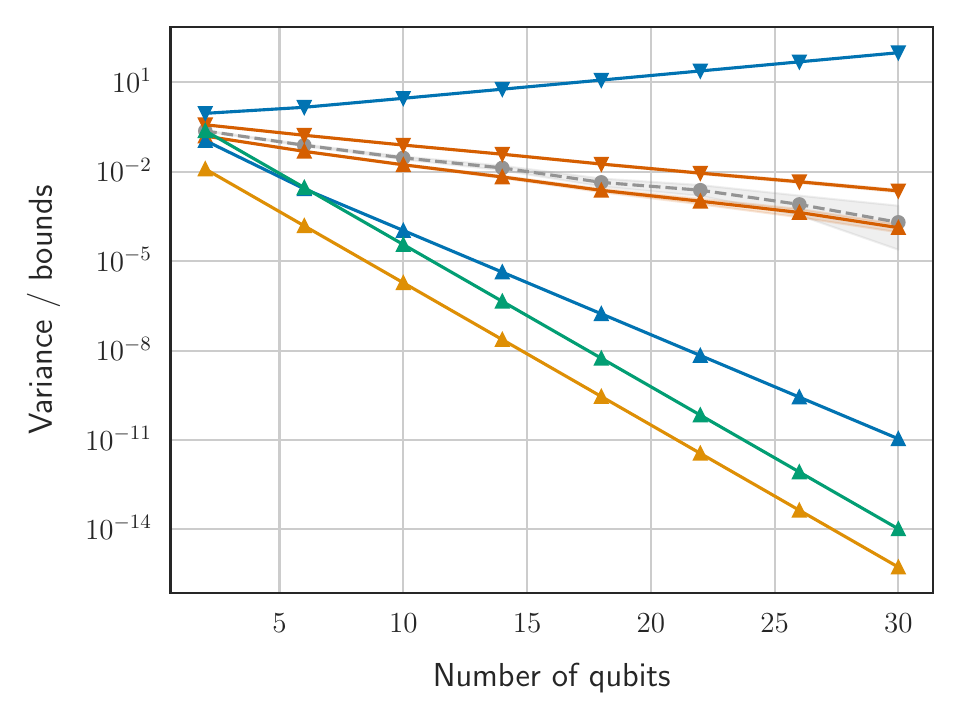}
\end{subfigure} \hfill
\caption{Variance of the loss function and corresponding bounds from Theorem \ref{th:bounds}, compared with existing work \cite{Cerezo_2021_costfunct, napp2022quantifying, Uvarov_2021BPs}, for two ansätze of depth $d = n/2$ and a 1-local observable. (a): EfficientSU2 ansatz. (b): Cartan ansatz forming an approximate local 2-design \cite{Uvarov_2021BPs}. Variance and bounds are estimated with $10000$ samples of $\T \sim \cD$. Full details, and further experiments with different circuit depths / observable locality, in Appendix \ref{app:efficient_evaluation} (Figure~\ref{fig:light-cones-bounds-appendix}).}
\label{fig:light-cones-bounds}
\end{figure*}

Notably, our reduction of a continuous variance to a discrete expectation implies that one can efficiently, and classically, estimate the variance of the loss, since $U(\T)$ is a Clifford circuit for parameters in $\{0, \pi/2\}$. This implies, in turn, that one can classically check whether a given circuit-observable pair will suffer from barren plateaus, hence enabling \textbf{classical ansatz design}.
Our bounds can similarly be estimated efficiently, and are significantly tighter than those of existing work, as illustrated in Figure~\ref{fig:light-cones-bounds}. We provide further experiments with different circuit depths and observable locality in Appendix~\ref{app:efficient_evaluation}, where we also show that the capability to efficiently evaluate $\Var_\T \big[ \cL_\al \big]$ does not make the bound evaluation redundant, because the variance of the bound estimation scales favorably compared with that of the loss.

An immediate consequence of Theorem \ref{th:bounds} is that we can rule out barren plateaus provided \textit{at least one} term $P_\al$ with non-vanishing $c_\al$ has logarithmically small light-cones. Given any mixed (or local) observable, one generic approach to achieve this is to construct a circuit with logarithmic depth and a local form of entanglement, since light-cones typically grow with each layer of entangling gates.
To make this as concrete as possible, we explicitly rule out barren plateaus for the standard class of EfficientSU2 circuits \cite{efficientsu2} in Corollary \ref{cor:su2} below. It should be noted that while this particular circuit may be simulated efficiently for a local observable, this is not necessarily possible in the mixed observable case. Furthermore, the setting can easily be extended to a larger class of circuits with shallow depth and $O(1)$-local entanglement, by varying the proof in Appendix \ref{app:bounds_proof}.

\begin{corollary}\label{cor:su2}
Let $H$ be any mixed observable on $n$ qubits, $\rho$ the zero initial state, and $U(\T)$ an EfficientSU2 circuit of logarithmic depth $O(\log n)$, pairwise entanglement, and rotation layers $(R_Y, R_Z)$. Then the corresponding loss does not suffer from barren plateaus, namely,
\[ \Var_\T \big[ \cL \big] \in \Omega\left( \frac{1}{\poly (n)}\right) \,. \]
\end{corollary}

For completeness, this corollary is illustrated numerically in Figure \ref{fig:locality_variance}\textcolor{red!80!black}{b}, where we provide explicit examples of local and mixed observables which induce only polynomially vanishing gradients -- as opposed to the exponentially vanishing gradients of a global observable.  Though the example is simplistic, the fact that mixed observables do not induce barren plateaus has a number of applications, and in particular, allows us to guarantee non-vanishing initial gradients for qGANs in Section \ref{sec:qgans}. It is also worth emphasizing that while global gradients vanish exponentially \textit{on average}, this does not imply we ought to use local  approximations of the loss, because global terms can contribute significantly later in training, as displayed in the training of qGANs (Fig.~\ref{fig:results}\textcolor{red!80!black}{a}).\\

\noindent\mypar{Extensions.} Corollary \ref{cor:su2} is stated for algebraically mixed observables, but can easily be extended with appropriate variations of the proof. We state two such possibilities below, which are in line with the corollaries 1 and 4 of \cite{napp2022quantifying}, proven for circuits forming 2-designs.

\begin{enumerate}
\item For 1D-lattice circuits, if $H$ contains a \textit{topologically} $k$-local Pauli string with $k \in O(\log n)$, then the lower bound is still polynomial. (Recall that an observable is topologically $k$-local if it acts non-trivially on $k$ \textit{neighbouring} qubits in the 1D lattice.)
\item If $H$ is \textit{low-bodied} \cite{rudolph2023trainability}, i.e. contains \mbox{an \textit{algebraically}} $k$-local Pauli string with $k \in O(\log n)$, then the lower bound is super-polynomial but sub-exponential, namely, $\Var_\T \big[ \cL \big] \in  \Omega(1/\poly(n^{\log n}))$.
\end{enumerate}

\noindent\mypar{Globality.}
As already mentioned, Theorem \ref{th:bounds} implies that even a \textit{global} observable may not necessarily induce barren plateaus. Although apparently at odds with \cite{Cerezo_2021_costfunct}, this is no real surprise, since global and local observables are only separated by a change of basis. Consider for example a circuit with two initial rotation layers, followed by a single linear layer of CNOT gates. The global observable $X^{\otimes{n}}$ propagates through this entanglement layer by cancelling itself out on all qubit lines except the last, producing a light-cone of size 1 and a gradient variance lower-bounded by $1/4$, \textit{for any number of qubits}. Despite the artificiality of this construction, it may be helpful in designing ansätze for problem-specific observables, and demonstrates that relaxing $t$-design assumptions allows us to capture finer observable-circuit interaction. Notably, this aligns with the recent insights on barren plateaus using symmetry groups and dynamical Lie algebras; see \cite{fontana2023adjoint, ragone2023unified, schatzki2022theoretical}.

In the following section, we apply Theorem \ref{th:bounds} to prove that a class of qGANs do not suffer from barren plateaus regardless of discriminator depth, providing a promising test-bed for generative quantum machine learning.

\subsection{Quantum GANs}\label{sec:qgans}

Generative QML models learn to encode the probability distribution underlying given data by employing quantum resources, which has been shown to facilitate efficient (but approximate) quantum data loading~\cite{Zoufal2019qGANs, Stamatopoulos2020optionpricingusing}. However, generative quantum models ~\cite{CoyleQBornSupremacy20, BenedettiGenModellin19, VarQBMZoufal20, QBMAmin18, QBMWiebe17}
that are based on \emph{explicit} loss functions such as the Kullback-Leibler divergence~\cite{kullback1951}--which require explicit access to the model  probability distribution--inherently suffer from observable-induced barren plateaus~\cite{Supanut21Subtleties, rudolph2023trainability}.
It is therefore desirable to work with \emph{implicit} loss functions, which are functions of samples underlying the model probability distribution, instead.
For example, \cite{rudolph2023trainability} have suggested the use of implicit loss functions which are approximately local, such as the Maximum Mean Discrepancy loss with a particular class of Gaussian kernels.
However, the trainability of models with this type of loss function has so far only been proven for tensor product ansätze.
Orthogonally, local formulations and approximations of the actual loss functions have been suggested as potential remedy \cite{BenedettiF-Divergences21, rudolph2023trainability, sharma2020trainabilityDissipative, bravoprieto2020variational, Volkoff_2021_LargeGradients, Khatri2019quantumassisted, HolmesVariationalFastForwarding20}. Although this approach can guarantee non-vanishing gradients, it may however lead to spurious local minima \cite{AnschuetzQuantumTraps22, CerezoVariationalQSE22}. 

On the other hand, hybrid qGANs \cite{Zoufal2019qGANs, killoran2018qgans, lloyd2qGANs18, romero21qGAN}, also based on implicit loss functions, can be reformulated as a ground state problem with respect to a global observable. We focus on this setting, which falls outside the scope of trainability guarantees proven by existing work.

A generative adversarial network consists of two opposing models, a generator and a discriminator. The goal of the generator is to generate data samples that are similar to a training data set and the goal of the discriminator is to correctly classify data samples as true (from the training data set) or fake (stemming from the generator).
We consider hybrid qGANs where the generator is a parameterized \textbf{quantum} circuit $G_\T$, while the discriminator is a \textbf{classical} neural network $D_\phi$ -- resulting in a hybrid qGAN. The generator's goal is to encode a \emph{discrete} distribution $p_\T$ matching the original distribution $p_\text{true}$ underlying the training data, while the discriminator attempts to distinguish between them.

\begin{definition}[Discriminator Class]\label{def:discr}
We consider discriminators $D_\phi : \B^n \to \R$  in the class $\cN$ of fully-connected neural networks with $L$ hidden layers, leaky-ReLU activation functions \cite{maas2013leaky} and an output activation $F(x) = \log(\sig(x))$, for min-max GANs, or $F(x) = x$, for Wasserstein GANs. For each layer $l$, we write $m_l$ for the number of neurons (width), $\sig_l$ for the standard deviation of initial weights (all parameters except biases), and $\ga_l$ for the leaky-ReLU parameter. In particular, the neural net has $m_0 = n$ inputs and $m_{L+1} = 1$ output.
We assume that discriminator weights and biases are initialized following any i.i.d.~symmetric distributions, such as state-of-the-art Kaiming or Xavier initializations \cite{UnderstandingDifficultyTrainingGlorot2010, he2015delvingRectifiers}. If the training data $x \in X$ is not initially binary, we assume that it can be transformed by some mapping $T : X \to \B^n$ before being fed into the discriminator.
\end{definition}

Importantly, note that that leaky-ReLU is the typical choice of activation for state-of-the-art GANs \cite{radford2015gans, brock2018gans, karras2020gans}.

\begin{definition}[Generator Class]\label{def:gen}
We consider generators $G_\T$ in the class $\cU$ of parameterized quantum circuits from Definition \ref{def:circuit}. The generator action is written as
\[ G_\T \ket{0}^{\otimes n}= \sum_{x \in \B^n} \sqrt{p_\T(x)} e^{i q_\T(x)} \ket{x} \,, \] 
noting that the phase factor $e^{i q_\T(x)}$ has no impact on the learnt distribution, and can therefore be neglected.
\end{definition}

\noindent The generator and discriminator losses are given by
\begin{align*}
\cL_G(\T, \phi) &= \E_{x \sim p_\T} \big[ F(D_\phi(x)) \big] \,, \\
\cL_D(\T, \phi) &= \E_{x \sim p_\text{true}} \big[ F(D_\phi(x)) \big] + \E_{x \sim p_\T} \big[ \tilde{F}(D_\phi(x)) \big]\,,
\end{align*}
for some analytic functions $F, \tilde{F} : \R \to \R$, encompassing both min-max \cite{goodfellow_2014_gans} and Wasserstein \cite{arjovsky2017wassersteingan} GANs. While the (classical) discriminator gradients may be computed with standard automatic differentiation techniques, the (quantum) generator gradients require further considerations.
The latter reads $\nabla_{\T}\cL_G(\T, \phi) =\sum_x  \nabla_{\T}p_\T(x)F(D_\phi(x))\,,$
where the individual gradients $\nabla_{\T}p_\T(x)$ are subject to observable-induced barren plateaus \cite{Cerezo_2021_costfunct} because $\nabla_{\T}p_\T(x) = \nabla_{\T} \bra{0} G_\T^\dagger O G_\T \ket{0}$, where $O=\ket{x}\bra{x}$ is a global projector.

However, this intuition is misleading: the generator loss can be rewritten as $\cL_G(\T, \phi) = \bra{0} G_\T^\dagger H_\phi G_\T \ket{0}$, with a diagonal observable $H_\phi = \sum_{x} F(D_\phi(x)) \ket{x}\bra{x}$ that depends only on the discriminator, and can be expanded into the $Z$-Pauli basis to obtain $H_\phi = \sum_\al c_\al(\phi) Z_\al$, where, using Equation \eqref{eq:weights},
\begin{equation}\label{eq:qgan_coeffs}
c_\al(\phi) = \left( \frac{1}{2^n} \sum_x {(-1)}^{\al \cdot x} D_\phi(x) \right) \,.
\end{equation}
Since these weights are independent from $\T$, we can moreover split the loss into a weighted sum that includes local and global terms $\cL_\al(\T) = \bra{0} G_\T^\dagger Z_\al G_\T \ket{0}$ which are, in turn, independent from $\phi$:
\begin{equation}\label{eq:qgan_loss}
\cL_G(\T, \phi) = \sum_\al c_\al(\phi) \cL_\al(\T) \,.
\end{equation}

Hence, following Theorem \ref{th:bounds}, $H_\phi$ will not necessarily induce a barren plateau if local weights $c_\al(\phi)$ are sufficiently large for local contributions to the gradient to be non-vanishing. The central contribution of this section, Theorem \ref{th:qgan}, is precisely to prove that this holds for all 1-local weights, for any discriminator from Definition \ref{def:discr}.

\begin{theorem}\label{th:qgan}
Consider any discriminator $\cD_\phi \in \cN$, and the corresponding weights $c_\al(\phi)$ defined by Equation \eqref{eq:qgan_coeffs}. Then, for any 1-local weight (namely $\abs{\al} = 1$), we have
\begin{align*}
\E_\phi \Big[ c_\al(\phi)^2 \Big] &\geq \frac{\sig_{L+1}^2}{16} \ \prod_{l=1}^{L} \frac{m_l \sig_l^2(1+\gamma_l)^2}{4} \,.
\end{align*}
In particular, initialising parameters such that $m_l \sig^2_l \geq 4$ for each $l$, the bound reduces to $\E_\phi \big[ c_\al(\phi)^2 \big] \geq \sig^2_{L+1}/16$, which is \textbf{constant} both in the number of qubits $n$ and the discriminator depth $L$.
\end{theorem}

Theorem \ref{th:qgan} guarantees that the observable $H_\phi$ induced by a discriminator $D_\phi \in \cD$ has 1-local weights that stay \textit{constant} in the number of qubits, provided the initial variance of parameters is inversely proportional to layer width. This coincides with the state-of-the-art approaches to initialisation in classical ML, guaranteeing that back-propagated gradients have a maintained variance across layers \cite{UnderstandingDifficultyTrainingGlorot2010, he2015delvingRectifiers}. Combining these insights with our previous results, we can immediately rule out barren plateaus for qGANs, provided the generator has logarithmic depth. We state this explicitly for the class of EfficientSU2 circuits in Corollary \ref{cor:qgan} below, which, as already discussed, can easily be extended to other circuit classes with shallow depth and $O(1)$-local entangling layers.

\begin{corollary}[Absence of barren plateaus in qGANs]\label{cor:qgan}
Let $\rho$ the zero initial state, $G_\T$ be an EfficientSU2 circuit of logarithmic depth $O(\log n)$, pairwise entanglement, and rotation layers $(R_Y, R_Z)$, and $D_\phi \in \cN$ be a discriminator satisfying $m_l \sig_l^2 \geq 4$ for each layer $l$. Then the induced loss function given by Equation \eqref{eq:qgan_loss} does not suffer from barren plateaus, namely,
\[ \Var_{\T, \phi} \big[ \cL_G \big] \in \Omega\left( \frac{1}{\poly(n)} \right) \,. \]
\end{corollary}

This guarantees that initialising parameters following Definitions \ref{def:discr} and \ref{def:gen}, namely, uniformly for the generator and following any i.i.d. symmetric distribution for the discriminator, produces non-exponentially-vanishing loss and gradients \cite{Arrasmith_2022}. One might also ask whether the (classical) discriminator has non-vanishing gradients -- but this has extensively been studied in classical machine learning, and using a leaky-ReLU activation function is known to mitigate the problem significantly \cite{glorot2011relu}.

As already emphasized, note that the absence of barren plateaus is a necessary condition for trainability, but in no way guarantees that training will lead to an optimal solution. Nonetheless, the fact that Pauli terms contribute independently (Theorem \ref{th:bounds}) proves that local approximations of the loss function can only harm the training process -- by increasing initial concentration of gradients and potentially leading to undesirable local minima. Moreover, while initially exponentially small, global terms may kick in over the course of training, as displayed numerically in the following section.

\subsection{Experiments}\label{sec:qgans-experiments}
\begin{figure*}[t!]
\centering
\begin{subfigure}[t]{0.49\linewidth}
\subcaption{Relative Entropy \& Gradients}
\includegraphics[width=\linewidth]{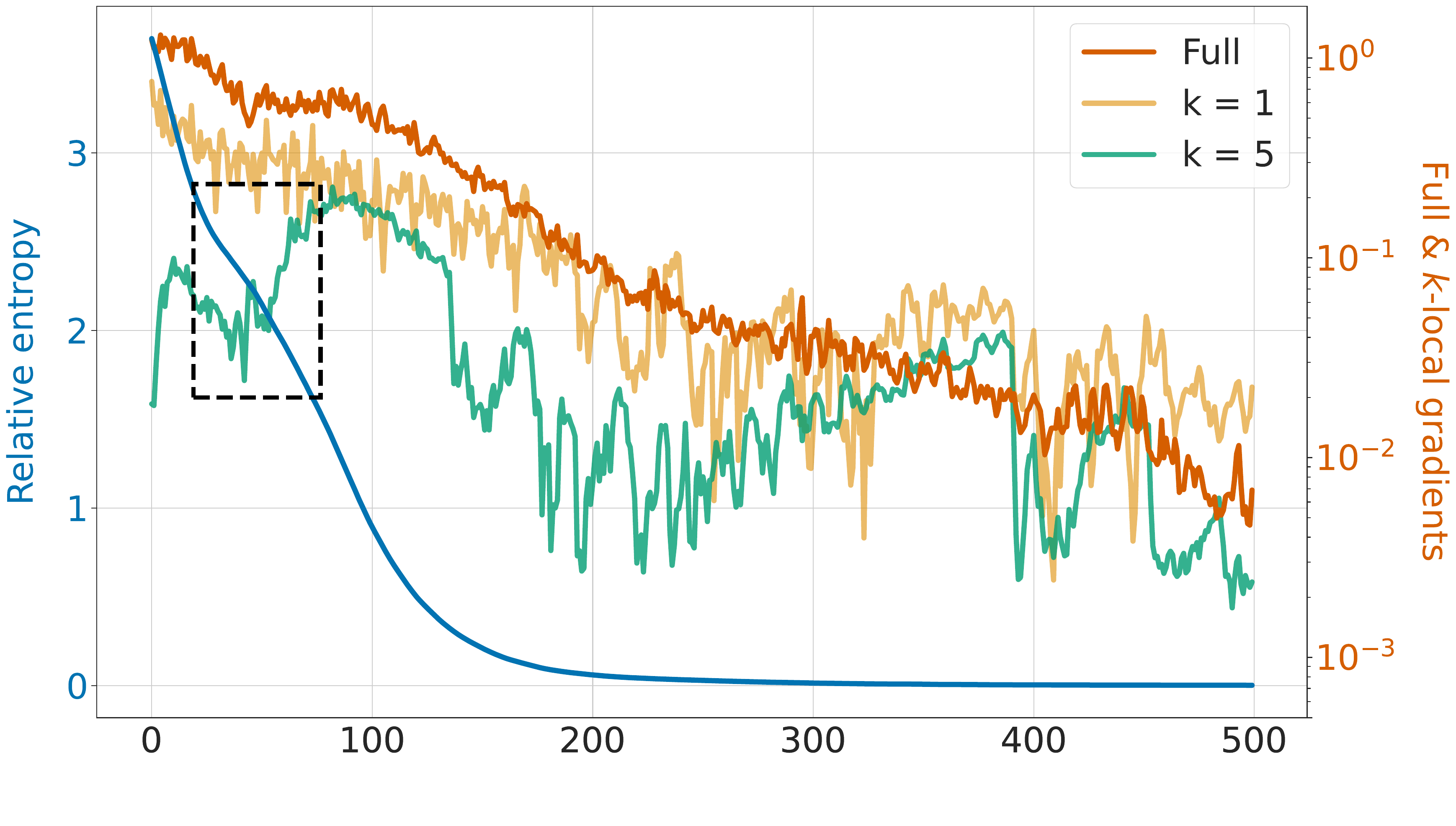} 
\end{subfigure} \hfill
\begin{subfigure}[t]{0.48\linewidth}
\subcaption{True \& Generated PDFs}\vspace{10pt}
\includegraphics[width=\linewidth]{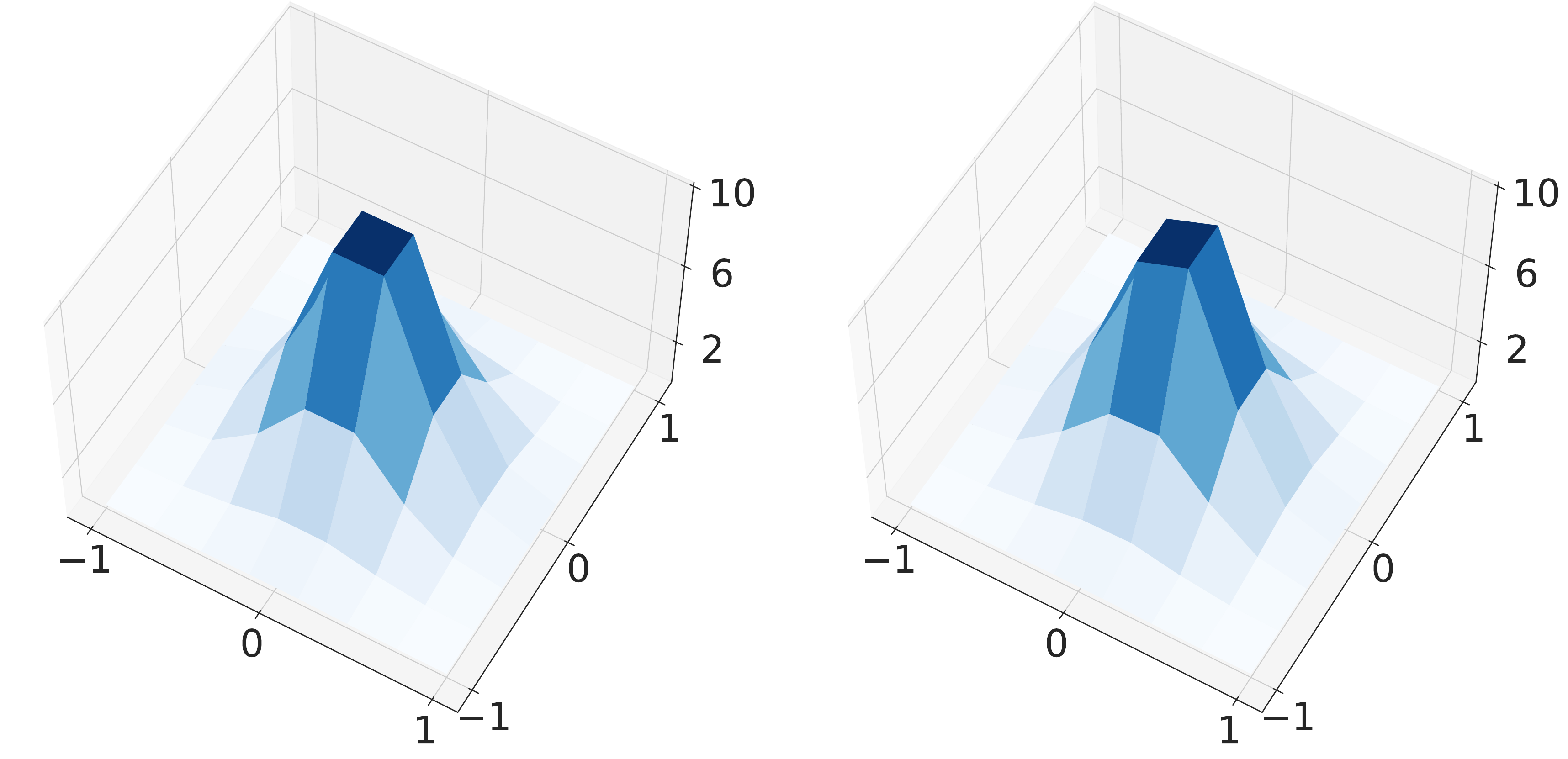}
\end{subfigure} \hfill
\begin{subfigure}[c]{0.49\linewidth}
\includegraphics[width=\linewidth]{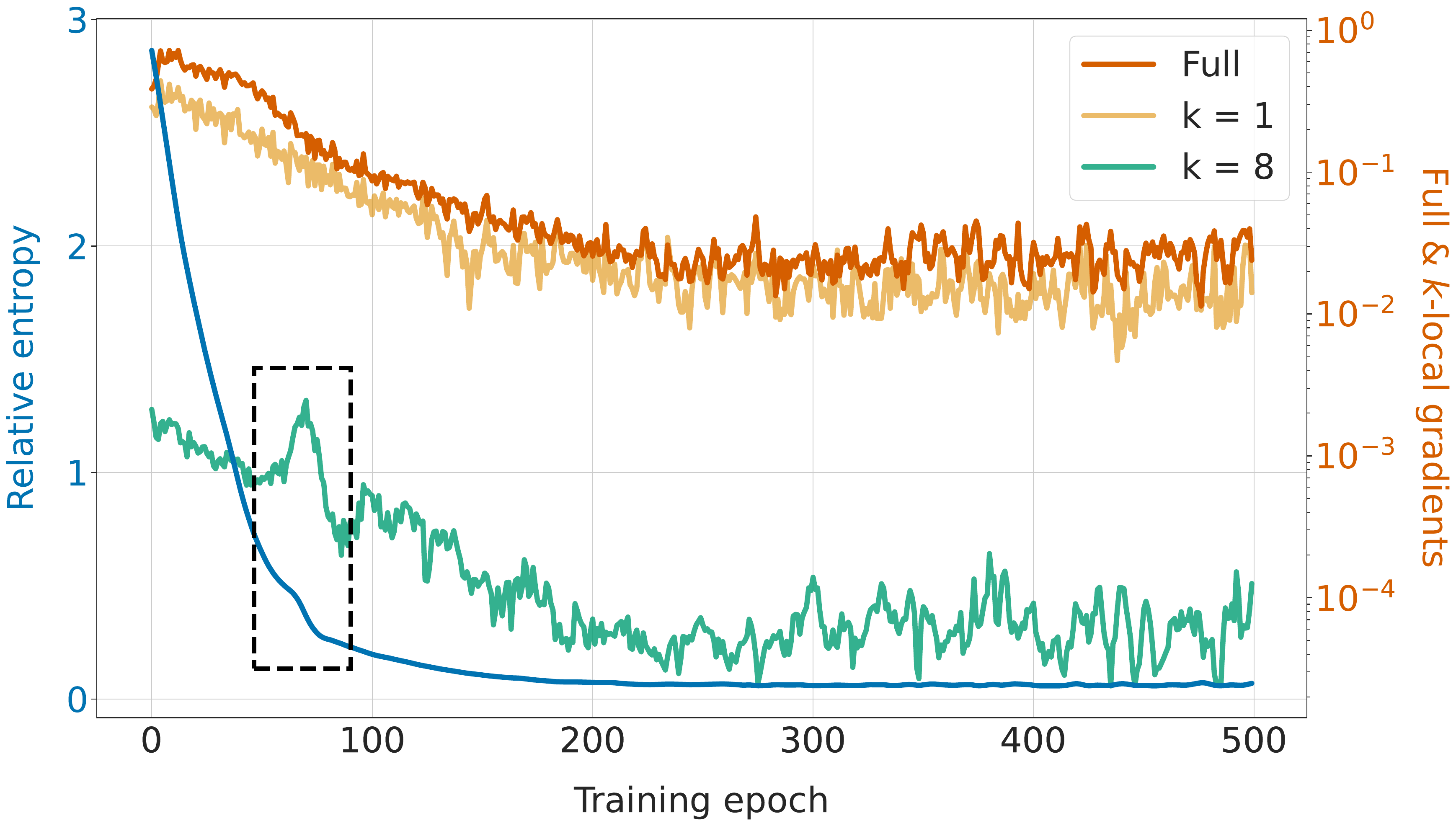} 
\end{subfigure} \hfill
\begin{subfigure}[c]{0.48\linewidth}
\includegraphics[width=\linewidth]{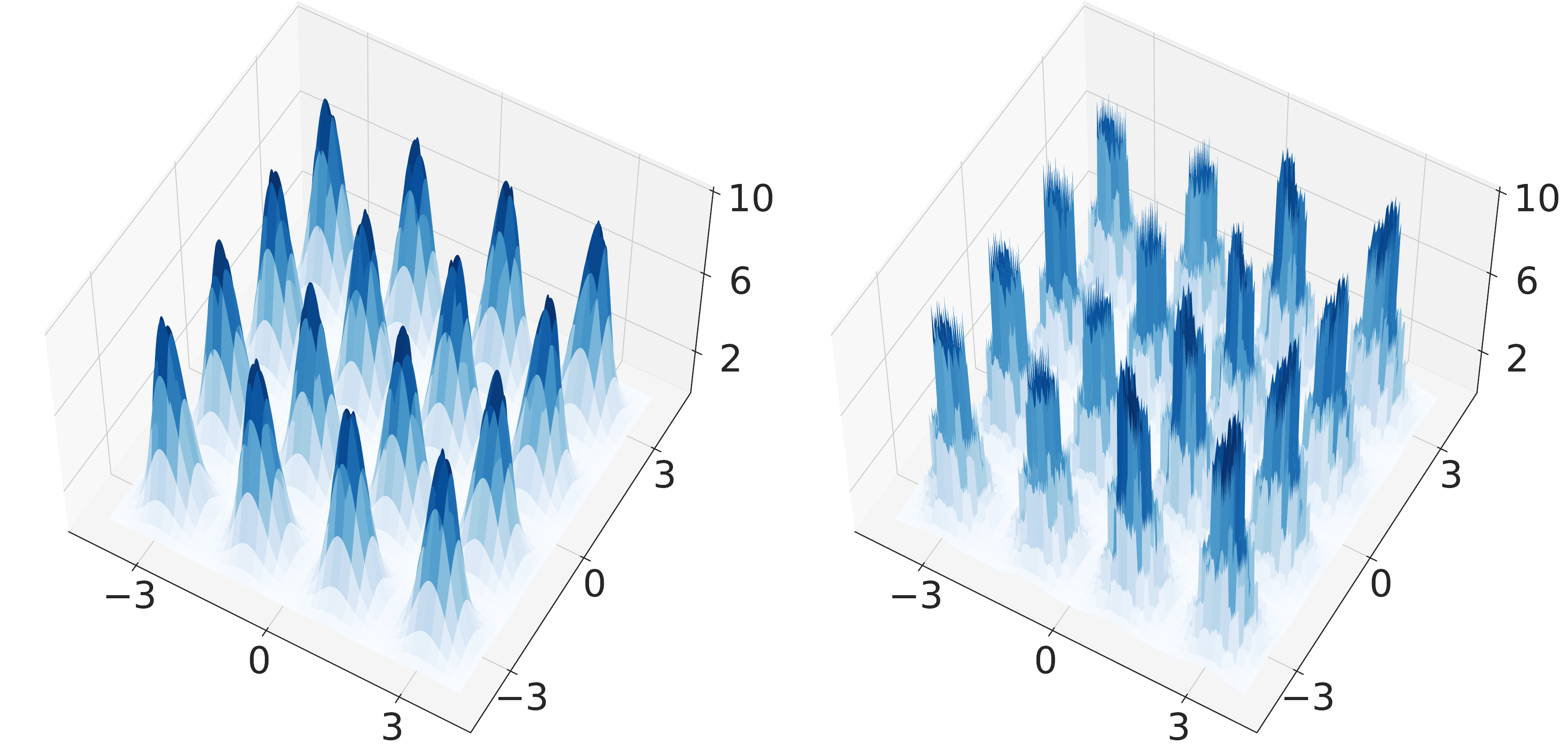} 
\end{subfigure} \hfill
\caption{Results for 6-qubit (top) and 16-qubit (bottom) experiments. (a): Relative entropy between true and generated distributions (blue curve), and 1-norm of generator gradients, over the course of training. The \textbf{full} gradient is simply the gradient of the generator loss $\cL_\cG$, while each \textbf{$k$-local} gradient is the gradient of $\sum_{\abs{\al} = k} c_\al \cL_\al$ from Equation \eqref{eq:qgan_loss}. Dashed boxes correspond to non-local gradients kicking in, following a period of deceleration in the relative entropy. (b): True (left) and generated (right) probability density functions at the end of training.}
\label{fig:results}
\end{figure*}

\subsub{Setup} To assess the relevance of our theoretical results, we reproduce a GAN experiment from \cite{letcher2019stable} corresponding to a simple 2D Gaussian distribution (with $n = 6$ qubits), and a difficult 2D \textit{mixture} of Gaussian distributions (with $n = 16$ qubits) by training a qGAN using Qiskit \cite{qiskit}. The mixture is a classically difficult distribution to learn, where optimization methods typically suffer from \textit{mode collapse} \cite{metz2016unrolled} and \textit{mode hopping} \cite{letcher2019game}. 

The generator is an EfficientSU2 circuit with two layers of single-qubit $R_Y$ and $R_X$ rotations, separated by a single layer of CNOT pairwise entanglement, satisfying the requirements of Corollary \ref{cor:qgan}. The parameters $\T$ are initialized uniformly over $[-\pi, \pi]$. The discriminator is a fully-connected neural network with $L = 3$ hidden layers of width $m = 64$ each, with leaky-ReLU hidden activation. For discriminator parameters $\phi$, we use a slight modification of Pytorch's \cite{pytorch} default initialization scheme, Kaiming uniform \cite{he2015delvingRectifiers}, with \texttt{fanmode = fanout} and twice the standard deviation. This guarantees constant 1-local weights following Theorem \ref{th:qgan}, while being in line with state-of-the-art initialization schemes for GANs. Both generator and discriminator parameters are optimized using Adam \cite{Kingmaadam14}, with learning rate $\al = 0.01$ and momentum terms $\be = (0.7, 0.999)$. The real and generated batch sizes are 256; the number of discriminator updates per generator update is $5$.
To enable a clear presentation, the results in this section are presented for a single seed of initial parameters and with exact gradients. To provide evidence of statistical significance and robustness of the estimated gradients, we provide additional experiments in Appendix \ref{app:averaged_experiments}.

\subsub{Results} The relative entropy between true and generated probability functions, as well as the generator gradients over the course of training, are shown in Figure \ref{fig:results}(a). In part (b), we plot the generator PDF (left) at the end of training, which can be seen to match the true PDF (right) quasi-perfectly for 6 qubits, and reasonably well for 16 qubits. In both experiments, the norm of the full gradient at initialization is seen to be large -- in line with our theoretical guarantees (Corollary \ref{cor:qgan}). The relative entropy consistently decreases over the course of training, and stabilizes towards 300 epochs. Interestingly, our results are on par with the results achieved using classical GANs in \cite{letcher2019stable} despite a massively smaller number of parameters: 64 parameters for our 16-qubit quantum generator, reaching a relative entropy of \tildenormal 0.06 in 300 training epochs, compared with 764930 parameters for the classical network used by \cite{letcher2019stable}, reaching a relative entropy of \tildenormal 0.04 in 10000 training epochs, with batch sizes of 256 in both cases. 
This is in line with previous empirical results \cite{Abbas_2021PowerofQNNs} which indicated beneficial capacity properties and faster training convergence with parameterized quantum models compared to comparable classical neural networks.
However, it remains to check whether this performance can still be achieved in the presence of noise from physical quantum hardware, and to see how wall-clock times compare.

\subsub{Global Contributions}

In order to investigate whether global terms of the observable contribute significantly over the course of training, Figure \ref{fig:results}(a) includes the norm of the gradients associated with different $k$-local terms, with $k = 1,5$ in the 6-qubit case and $k=1,8$ in the 16-qubit case. One might also wish to plot the coefficients of $k$-local terms over the course of training, but we found these to be extremely highly correlated with the magnitude of corresponding gradients, adding no information value.

We notice that the global terms are initially much smaller than the 1-local terms, in agreement with theoretical results, but kick in non-trivially towards the 100th epoch. As highlighted in the dashed boxes, this seems to coincide with the entropy stagnating a little, and then being pushed further down, possibly thanks to this global contribution.
This observation was reproduced across multiple seeds, and suggests that global terms are central in reaching desirable local minima -- instead of discarding them, as suggested in prior work \cite{BenedettiF-Divergences21, bravoprieto2020variational, Volkoff_2021_LargeGradients, Khatri2019quantumassisted, rudolph2023trainability}.  

\subsub{Limitations} It is useful to notice that our 16-qubit experiments consistently led to `spiky' generator distributions, as shown in Figure \ref{fig:results}(b). Indeed, while the ground distribution is a \textit{continuous} Gaussian mixture, the generator encodes this into amplitudes $p_\T(x)$, where $x$ ranges over a \textit{discrete} set $\{0, 1\}^n$. Without introducing an inductive bias that forces `adjacent' inputs $x$ to have similar amplitudes, the generator is unfortunately blind to the underlying continuity, attempting to match the Gaussian mixture as if it were discrete. This may explain the resulting spikiness, and calls for further work on inductive biases and the optimization landscape of qGANs more generally.
%

\section{Conclusion and Outlook}
\label{sec:conclusion}
This paper extends prior work on barren plateaus, by lifting $t$-design assumptions and providing tight upper and lower bounds on gradient concentration for a large class of parameterized quantum circuits, and arbitrary observables (Theorem \ref{th:bounds}). Notably, we also present a method for classically estimating these bounds and the actual loss (gradient) variance efficiently, providing a practical tool for studying variational loss landscapes and designing ansätze.
In particular, our result emphasizes that gradient concentration is not exclusively determined by observable locality, circuit depth, and entanglement, but instead, strongly relies on the observable-ansatz \textit{interaction}, formulated in terms of light-cones. One fortunate consequence is that \textit{mixed} observables do \textit{not} necessarily induce barren plateaus (Corollary \ref{cor:su2}), and even \textit{global} observables may have non-vanishing gradients if the circuit is chosen wisely.
Interesting directions for future work would be to extend our proof techniques to additional circuit classes, such as Hamiltonian variational ansätze, as well as investigate the simulability of our circuit class, following \cite{cerezo2024doesprovableabsencebarren}.

Moreover, we leverage our results to show that qGANs are an auspicious class of GQML algorithms by proving that the corresponding observable is mixed under reasonable conditions, and hence, does not induce barren plateaus (Theorem \ref{th:qgan} and Corollary \ref{cor:qgan}).
We illustrate these theoretical results with numerical experiments, where we train a qGAN to learn Gaussian mixtures with up to $16$ qubits, and provide evidence that global contributions kick in during training.
We nonetheless emphasise that non-vanishing gradients, though necessary, are not a sufficient condition in the over-arching goal of reaching a global (or even a local) minimum of the loss.
In particular, while our work rules out barren plateaus for qGANs with shallow generators, future work calls for a deeper understanding of VQA optimization landscapes, and the potential introduction of an inductive bias to help qGANs efficiently learn continuous distributions, as suggested by our $16$-qubit experiments.

Finally, it is worth mentioning that our results highlight a central difference between quantum barren plateaus and the classical vanishing gradient problem for neural networks, where it is only the \textit{depth} of the circuit which causes gradients to vanish, by composition of activation functions across layers \cite{glorot2011relu, basodi2020gradient}. Instead, Theorem \ref{th:bounds} shows that quantum gradient concentration depends on how many qubits are `hit' by the observable (the \textit{width} of the light-cone), and thus, whether the sphere on which they live induces a concentration of measure phenomenon \cite{Clean_2018_BarrenPlateaus}. Depth of the circuit, in itself, \textit{does not} induce vanishing gradients -- it only does so indirectly because each layer typically entangles more and more qubits together, thus broadening the light-cone. Another way of making this point is that high \textit{local} expressivity (high-depth on a local part of the circuit) does not induce barren plateaus -- whereas the analogous situation for classical neural networks \textit{would}.\\

\noindent\textbf{Data Availability.} The presented data can be made available upon reasonable request.

\noindent\textbf{Code Availability.} The code can be made available upon reasonable request.

\noindent\textbf{Acknowledgments.} We thank Francesco Tacchino and Alberto Baiardi for their input on quantum chemistry applications, Marco Cerezo for interesting discussions, and Kunal Sharma and Zo\"{e} Holmes for providing valuable feedback and input on the manuscript. SW acknowledges the support of the SNF grant No. 214919. SW and CZ acknowledge the support of the SNF grant No. 215933.

\bibliography{references}

\pagebreak
\onecolumngrid
\appendix
\pagebreak
\setlength{\parindent}{0pt}

\section{Proof of Theorem \ref{th:bounds}}\label{app:bounds_proof}

\begin{mybox}
\begin{apptheorem}{\ref{th:bounds}}
For any $m$-parameter circuit $U(\T) \in \cU$, any observable $H = \sum_\al c_\al P_\al$, and any mixed state $\rho$, each Pauli term makes an independent contribution to the loss and gradient variances:
\begin{align*}
\Var_\T \big[ \cL \big] &= \sum_\al c_\al^2 \Var_\T \big[ \cL_\al \big] \,, \\
\Var_\T \big[ \nabla \cL \big] &= \sum_\al c_\al^2 \Var_\T \big[ \nabla \cL_\al \big] \,.
\end{align*}
Moreover, writing $\cD$ for the discrete uniform distribution over $\{0, \pi/2\}^m$, each non-trivial term ($\al \neq 0$) can be reduced to a \textbf{discrete} expectation, namely,
\[ \norm{\, \Var_\T \big[ \nabla \cL_\al \big] \,}_{\infty} = \Var_\T \big[ \cL_\al \big] = \E_\cD \big[ \cL^2_\al \big] \,. \]
Finally, for product mixed states $\rho$ (generalized in Part \textbf{(G)} of the proof), the loss variance is bounded by
\begin{gather*}
\Om(\rho) \, \E_{\cD} \left[ \  \left( \frac{1}{4} \right)^{\conealphatheta} \ \right] \leq \Var_\T \big[ \cL_\al \big] \leq \E_{\cD} \left[ \ \left( \frac{1}{2} \right)^{\conealphatheta} \ \right] \,,
\end{gather*}
and there are circuit-observable pairs that saturate these inequalities. In particular, combined with the previous equation, at least one partial derivative $\partial_k \cL_\al$ satisfies the lower bound, and all of them satisfy the upper bound.
\end{apptheorem}
\end{mybox}

\begin{proof} To avoid overloading indices, we make the following notational changes. We write
\[ P(\T) \coloneqq R_P(\T) = \exp(-iP\T/2) \]
for the rotation gate induced by a Pauli string $P$ and a parameter $\T$. By introducing identity Clifford gates $W_k$ in order to separate each parameterized rotation into its own layer $R_k$, and adjusting the value of $K$ accordingly, we rewrite the circuit with relabelled parameters $\bT = (\T, \phi, \omega)$ as
\begin{equation}
U(\bT) = A_K(\T) B(\TT) C(\TTT) \,, \label{eq:circuit_rewrite}
\end{equation}
where
\[ A_K(\T) = \prod_{k=1}^K W_k P_k(\T_k) \qquad , \qquad B(\TT) = \bigotimes_{i=1}^n \sig_{\mu_i}(\TT_i) \qquad , \qquad C(\TTT) = \bigotimes_{i=1}^n \sig_{\nu_i}(\TTT_i) \,, \]
so that $\sig_{\mu_i}, \sig_{\nu_i}$ are the two layers of orthogonal single-qubit rotations, i.e. $\mu, \nu \in \{1, 2, 3\}^n$ and $\mu_i \neq \nu_i$ for each $i$. We will write $\E$ without subscript to denote the expectation over all parameters $\E_{\bT}$. To begin, recall that decomposing the observable $H = \sum_\al c_\al P_\al$ leads to a decomposition of the loss
\[ \cL = \Tr\left(U^\dagger H U \rho \right) = \sum_\al c_\al \Tr\left( \rho U^\dagger P_\al U \right) \eqqcolon c_\al \cL_\al \,, \]
and correspondingly, a decomposition of partial derivatives $\partial_\tau \cL = c_\al \partial_\tau \cL_\al$ for each parameter $\tau \in \{\T, \TT, \TTT\}$. Finally, we also decompose the initial state $\rho = \sum_\la d_\la P_\la$ into the Pauli basis.\\

\mypar{(A) Vanishing expectation.} In order to prove that Pauli terms contribute independently to loss and gradient \textit{variances}, we first prove that loss and gradient \textit{expectations} vanish. To begin, by linearity, notice that the expectations
\begin{gather*}
\E \big[ \cL \big] = \sum_\al c_\al \E \big[ \cL_\al \big] \qquad \text{and} \qquad
\E \big[ \partial_\tau \cL \big] = \sum_\al c_\al \E \big[ \partial_\tau \cL_\al \big]
\end{gather*}
decompose across Pauli strings. For all $\al \neq 0$, we prove that $\E \big[ \cL_\al \big] = \E \big[ \partial_\tau \cL_\al \big] = 0$ by induction on the number of layers $K$. For the base case ($K = 0$), the loss splits across qubits (indexed by $i$) as follows:
\begin{align}\label{eq:loss_base}
\cL_\al = \sum_\la d_\la \Tr\left(U^\dagger P_\al U P_\la \right) = \sum_\la d_\la \prod_i \Tr\big(\sig_{\nu_i}(-\TTT_i) \sig_{\mu_i}(-\TT_i) \sig_{\al_i} \sig_{\mu_i}(\TT_i) \sig_{\nu_i}(\TTT_i) \sig_{\la_i} \big) \eqqcolon \sum_\la d_\la \prod_i \cL_{\al\la}^i \,.
\end{align}
By assumption that $\al \neq 0$, there exists $j$ such that $\al_j \neq 0$. If $\al_j \neq \mu_j$, we have
\begin{align*}
\cL_{\al\la}^j &= \Tr\big(\sig_{\nu_j}(-\TTT_j) \sig_{\mu_j}(-2\TT_j) \sig_{\al_j} \sig_{\nu_j}(\TTT_j) \sig_{\la_j} \big) \\
&= \cos(\TT_j) \Tr\big(\sig_{\nu_j}(-\TTT_j) \sig_{\al_j} \sig_{\nu_j}(\TTT_j) \sig_{\la_j} \big) + i \sin(\TT_j) \Tr\big(\sig_{\nu_j}(-\TTT_j) \sig_{\mu_j} \sig_{\al_j} \sig_{\nu_j}(\TTT_j) \sig_{\la_j} \big) \,.
\end{align*}
Both terms vanish on expectation over $\TT_j$. On the other hand, if $\al_j = \mu_j$, orthogonality of $\mu, \nu$ implies $\al_j \neq \nu_j$, hence
\begin{align*}
\cL_{\al\la}^j = \Tr\big(\sig_{\nu_j}(-2\TTT_j) \sig_{\al_j} \sig_{\la_j} \big) = \cos(\TTT_j) \Tr\big( \sig_{\al_j} \sig_{\la_j} \big) + i\sin(\TTT_j) \Tr\big( \sig_{\nu_j} \sig_{\al_j} \sig_{\la_j} \big)\,.
\end{align*}
This also vanishes on expectation over $\TTT_j$. By independence between parameters, we obtain $\E \left[ \cL_\al \right] = 0$, concluding the base case. For the induction step, writing $U_K$ for a circuit with $K \geq 1$ layers, we have
\begin{align*}
\cL_\al = \Tr\left(U_K^\dagger P_\al U_K  \rho \right) = \Tr\left(U_{K-1}^\dagger P_K(-\T_K) W_K^\dagger P_\al W_K P_K(\T_K) U_{K-1}  \rho \right) \,.
\end{align*}
Since $W_K$ is a Clifford gate, there exists a non-identity Pauli string $P_\ga$ such that $W_K^\dagger P_\al W_K = P_\ga$. Now note that $PQ = \pm QP$ for all Pauli strings $P, Q$, and therefore $P_K(\T) P_\ga = P_\ga P_K(\pm \T)$ depending on whether $P_K$ commutes with $P_\ga$. In the former case, we obtain
\begin{align*}
\E \left[ \cL_\al \right] = \Tr\left(U_{K-1}^\dagger P_\ga U_{K-1} \rho  \right) = 0
\end{align*}
by the inductive hypothesis. Otherwise, we have
\begin{align*}
\cL_\al = \cos \T_K \Tr\left( U_{K-1}^\dagger P_\ga U_{K-1}  \rho  \right) + i \sin \T_K \Tr\left( U_{K-1}^\dagger P_K P_\ga U_{K-1}  \rho  \right) \,,
\end{align*}
which vanishes on expectation over $\T_K$. This concludes the induction, so we obtain $\E \big[ \cL_\al \big] = 0$ for all $\al \neq 0$ and hence $\E \big[ \cL \big] = c_0$. For the gradient, we can similarly prove by induction that $\E \big[ \partial_\tau \cL_\al \big] = 0$ for all $\al$, but there is a simpler argument that has also been invoked in the work of \cite{Arrasmith_2022}: notice that $\cL$ is periodic in each parameter $\tau$ on $[-\pi, \pi]$, so the integral of $\partial_\tau \cL$ over $\tau$ must vanish, hence $\E \big[ \partial_\tau \cL \big] = \E \E_{\tau} \big[ \partial_\tau \cL \big] = 0$.\\

\noindent\mypar{(B) Vanishing covariance.} Second, we prove that covariances between Pauli strings also vanish, namely,
$\E \left[ \cL_\al \cL_\be \right] = \E \left[ \partial_\tau \cL_\al \partial_\tau \cL_\be \right] = 0$ for all $\al \neq \be$ and for all parameters $\tau$. To begin, let $j$ be any index such that $\al_j \neq \be_j$. For the base case, recall from equation \eqref{eq:loss_base} that $\cL_\al = \sum_\la d_\la \prod_i \cL_{\al\la}^i$ and hence, by parameter independence,
\[ \E \left[ \cL_\al \cL_\be \right] = \sum_{\la, \la'} d_\la d_{\la'} \prod_i \E \left[ \cL_{\al\la}^i \cL_{\be\la'}^i \right] \,. \]
It is thus sufficient to show that the $j$th term vanishes for all $\la, \la'$. If $\al_j = 0$ (without loss of generality), we have $\be_j \neq 0$, hence
\[ \E \left[ \cL_{\al\la}^j \cL_{\be\la'}^j \right] = \E \left[ \cL_{\be\la'}^j \right] = 0 \]
by section \textbf{(A)}. We can therefore assume that $\al_j, \be_j \neq 0$. Now if $\al_j = \mu_j$ (wlog), the first loss $\cL_{\al\la}^j$ is independent from $\TT_j$. On the other hand, we must have $\be_j \neq \mu_j$ and hence
\begin{align*}
\cL_{\be\la'}^j &= \cos(\TT_j) \Tr\big(\sig_{\nu_j}(-\TTT_j) \sig_{\be_j} \sig_{\nu_j}(\TTT_j) \sig_{\la'_j} \big) + i \sin(\TT_j) \Tr\big(\sig_{\nu_j}(-\TTT_j) \sig_{\mu_j} \sig_{\be_j} \sig_{\nu_j}(\TTT_j) \sig_{\la'_j} \big) \,,
\end{align*}
which vanishes on expectation over $\phi_j$. We therefore obtain 
\[ \E \left[ \cL_{\al\la}^j \cL_{\be\la'}^j \right] = \E \left[ \cL_{\al\la}^j \E_{\TT_j} \left[ \cL_{\be\la'}^j \right] \right] = 0 \,. \]
It remains only to deal with the final case, where $\al_j \neq \be_j \neq \mu_j$. By Lemma \ref{lem:trig}, integrating over $\TT_j$ yields
\begin{align*}
\E_{\TT_j} \left[ \cL_{\al\la}^j \cL_{\be\la'}^j \right] &= \frac{1}{2} \Tr\left( \sig_{\nu_j}(-\TTT_j) \sig_{\al_j} \sig_{\nu_j}(\TTT_j) \sig_{\la_j} \right) \Tr\left( \sig_{\nu_j}(-\TTT_j) \sig_{\be_j} \sig_{\nu_j}(\TTT_j) \sig_{\la'_j} \right) \\
&+ \frac{1}{2} \Tr\left( \sig_{\nu_j}(-\TTT_j) i \sig_{\mu_j} \sig_{\al_j} \sig_{\nu_j}(\TTT_j) \sig_{\la_j} \right) \Tr\left( \sig_{\nu_j}(-\TTT_j)  i \sig_{\mu_j} \sig_{\be_j} \sig_{\nu_j}(\TTT_j) \sig_{\la'_j} \right) \,.
\end{align*}
Now $\al_j \neq \be_j \neq \mu_j$ combined with $\mu_j \neq \nu_j$ implies (wlog) that $\al_j = \nu_j$ and $\be_j \neq \nu_j$. The first term thus reduces to
\[ \frac{1}{2} \Tr\left( \sig_{\al_j} \sig_{\la_j} \right) \Big( \cos(\TTT_j) \Tr\left( \sig_{\be_j} \sig_{\la'_j} \right) + i \sin(\TTT_j) \Tr\left( \sig_{\nu_j} \sig_{\be_j} \sig_{\la'_j} \right) \Big) \,, \]
which vanishes on expectation over $\TTT_j$. Similarly, the second term vanishes because $\al_j \neq \be_j \neq \mu_j$ combined with $\mu_j \neq \nu_j$ implies $i \sig_{\mu_j} \sig_{\al_j} \neq \pm \sig_{\nu_j}$ and $i \sig_{\mu_j} \sig_{\be_j} = \pm \sig_{\nu_j}$. We thus conclude $\E \left[ \cL_{\al\la}^j \cL_{\be\la'}^j \right] = 0$ for all $\la, \la'$ and hence $\E \left[ \cL_\al \cL_\be \right] = 0$, completing the base case. For the induction step, as in the previous section, let $U_K$ be a circuit with $K \geq 1$ layers and let $\ga, \del$ be such that $W_K^\dagger P_{\al/\be} W_K = P_{\ga/\del}$. If $P_K$ commutes with $P_\ga$ and $P_\del$, we obtain
\begin{align*}
\cL_\al \cL_\be = \Tr(U_{K-1}^\dagger P_\ga U_{K-1}\rho) \Tr(U_{K-1}^\dagger P_\del U_{K-1} \rho ) \,,
\end{align*}
which vanishes by the induction hypothesis. (Note that orthogonality of $P_\al, P_\be$ implies orthogonality of $P_\ga, P_\del$.) If $P_K$ commutes with $P_\ga$ but not with $P_\del$ (without loss of generality), we obtain
\begin{align*}
\cL_\al \cL_\be = \cos \T_K \Tr( U_{K-1}^\dagger P_\ga U_{K-1} \rho) \Tr(U_{K-1}^\dagger P_\del U_{K-1} \rho) + \sin \T_K \Tr(U_{K-1}^\dagger P_\ga U_{K-1} \rho) \Tr(U_{K-1}^\dagger i P_K P_\del U_{K-1}\rho) \,,
\end{align*}
which vanishes on expectation over $\T_K$. Finally, if $P_K$ commutes with neither, we obtain, by Lemma \ref{lem:trig},
\begin{align*}
\E_{\T_K} \left[ \cL_\al \cL_\be \right] &= \frac{1}{2} \Tr(U_{K-1}^\dagger P_\ga U_{K-1} \rho) \Tr( U_{K-1}^\dagger P_\del U_{K-1} \rho) \\
&+ \frac{1}{2} \Tr(U_{K-1}^\dagger iP_K P_\ga U_{K-1} \rho) \Tr(U_{K-1}^\dagger i P_K P_\del U_{K-1} \rho) \,.
\end{align*}
Since $P_\ga$ and $P_\del$ are orthogonal, $i P_K P_\ga$ and $i P_K P_\del$ are also orthogonal, so both terms vanish by the induction hypothesis. This completes the induction, giving $\E \left[ \cL_\al \cL_\be \right] = 0$. The proof for partial derivatives similarly follows by induction, or can be obtained using the parameter-shift rule
\[ \partial_\tau \cL(\bT) = \frac{\cL(\bT + e_\tau \pi/2) - \cL(\bT - e_\tau \pi/2)}{2} \,, \]
where $e_\tau$ is the unit vector along the $\tau$ direction, along with the standard trigonometric identities identities
\[ \cos(x \pm \pi/2) = \mp \sin(x) \qquad \text{and} \qquad \sin(x \pm \pi/2) = \pm \cos(x) \,. \]
From the vanishing expectations obtained in part \textbf{(A)}, we can now decompose the variance of the loss as
\[ \Var \big[ \cL \big] = \E \Big[ \left( \cL - c_0 \right)^2 \Big] = \E \left[ \bigg( \sum_{\al \neq 0} c_\al \cL_\al \bigg)^2 \right] = \sum_{\al, \be \neq 0} c_\al c_\be \E \left[ \cL_\al \cL_\be \right] = \sum_{\al \neq 0} c_\al^2 \E \left[ \cL_\al^2 \right] =  \sum_{\al} c_\al^2 \Var \big[ \cL_\al \big] \,. \]
Similarly, for gradients, we have, for each parameter $\tau$,
\[ \Var \big[ \partial_\tau \cL \big] = \sum_{\al} c_\al^2 \E \left[ \left( \partial_\tau \cL_\al \right)^2 \right] = \sum_{\al} c_\al^2 \Var \big[ \partial_\tau \cL_\al \big] \]
In other words, we have proven that each Pauli term $P_\al$ contributes independently to loss and gradient variances.\\

\mypar{(C) Variance of the gradients.} Note that a lower bound could be produced by invoking the converse of \cite[Theorem 1]{Arrasmith_2022}, but the bound would be loose. Instead, we directly show that
\[  \norm{\, \Var_\T \big[ \nabla \cL_\al \big] \,}_{\infty} \coloneqq \max_\tau \Var \big[ \partial_\tau \cL_\al \big] = \Var \big[ \cL_\al \big] \,. \]
For convenience, and by the equivalence obtained in part \textbf{(B)} above, we work with squared-expectations rather than variances. To begin, for any parameter $\tau$, we can invoke the parameter-shift rule to obtain
\begin{align*}
\E \left[ \left(\partial_\tau \cL_\al \right)^2 \right] &= \frac{1}{4} \Big( \E \left[ \cL_\al(\bT+e_\tau \pi/2)^2 \right] + \E \left[ \cL_\al(\bT-e_\tau \pi/2)^2 \right] - 2 \E \left[ \cL_\al(\bT+e_\tau \pi/2)\cL_\al(\bT-e_\tau \pi/2) \right] \Big) \,.
\end{align*} 
The first two terms can be equated with $\E \left[ \cL_\al(\bT)^2 \right]$ by changing variables $\bT \to \bT \pm e_\tau \pi/2$ and recalling that $\cL$ is periodic over each parameter with period $2\pi$, implying invariance over the uniform distribution. The third term can also be upper-bounded by $\E \left[ \cL_\al(\bT)^2 \right]$ by following the same induction leading to Equation \eqref{eq:induction_T}. We thus obtain
\begin{align*}
\E \left[ \left(\partial_\tau \cL_\al \right)^2 \right] &\leq \frac{1}{4} \Big( \E \left[ \cL_\al(\bT)^2 \right] + \E \left[ \cL_\al(\bT)^2 \right] + 2 \E \left[ \cL_\al(\bT)^2 \right]\Big) = \E \left[ \cL_\al(\bT)^2 \right]
\end{align*} 
for all $\tau$, and therefore $\max_\tau \E \left[ \left(\partial_\tau \cL_\al \right)^2 \right] \leq \E \left[ \cL_\al^2 \right]$. To prove equality, consider any parameter $\tau$ whose index we denote by $k$, and decompose $U = U_{\tau_+} U_{\tau_-}$, where $U_{\tau_-}$ and $U_{\tau_+}$ are the left- and right-hand sides of the parameterized gate $P_k(\tau)$, where $U_{\tau_-}$ includes the gate itself. In particular, partial derivatives can conveniently be written as
\[ \partial_\tau \cL_\al = \frac{i}{2} \Tr(U_{\tau_-}^\dagger \Big[ P_k, U_{\tau_+}^\dagger P_\al U_{\tau_+} \Big] U_{\tau_-} \rho) \,. \]
We now prove that equality is reached for the last (right-most) parameter $\tau$ for which $P_k(\T)$ does not commute with $U_{\tau_+}^\dagger P_\al U_{\tau_+}$. (For example, if the circuit ends with a parameterized $R_Z$ gate and $P_\al = X / Y$ on the corresponding qubit line, $\tau$ will be this final parameter. However, if $P_\al = I / Z$, then the final gate will have no impact and $\tau$ will be an earlier parameter.) Since all parameters beyond $P_k(\tau)$ are now defined to have no bearing on the loss, they can be set to $0$. By defining the Pauli string $P_\ga \coloneqq U_{\tau_+}^\dagger(0) P_\al U_{\tau_+}(0)$, thanks to each $W_k$ being Clifford, the loss reduces to
\[ \cL_\al = \Tr(U_{\tau_-}^\dagger P_\ga U_{\tau_-} \rho) \,. \]
Similarly, the partial derivative with respect to $\tau$ reduces to
\[ \partial_\tau \cL_\al = \frac{i}{2} \Tr(U_{\tau_-}^\dagger \Big[ P_k, P_\ga \Big] U_{\tau_-} \rho) \,. \]
Since $P_k$ is defined not to commute with $P_\ga$, we now have
\[ \frac{i}{2} \big[ P_k, P_\ga \big] = iP_k P_\ga = P_k(-\pi/2) P_\ga P_k(\pi/2) \,, \]
which implies that the partial derivative with respect to $\tau$ is simply a shifted version of the loss:
\[ \partial_\tau \cL_\al(\bT) = \Tr(U_{\tau_-}^\dagger(\bT-e_\tau \pi/2) P_\ga U_{\tau_-}(\bT+e_\tau \pi/2) \rho) = \cL_\al(\bT + e_\tau \pi/2) \,. \]
Finally, by changing variables $\bT \to \bT+e_\tau \pi/2$ and invoking the periodicity of $\cL$ once more, we obtain
\[ \E \big[ \big( \partial_\tau \cL_\al \big)^2 \big] = \E \big[ \cL_\al^2 \big] \]
for this particular parameter $\tau$, and therefore $\max_\tau \E \left[ \left(\partial_\tau \cL_\al \right)^2 \right] = \E \left[ \cL^2_\al \right]$. Rewriting this as a variance, we conclude
\[ \norm{\, \Var \big[ \nabla \cL_\al \big] \,}_{\infty} = \max_\tau \Var \big[ \partial_\tau \cL_\al \big] = \Var \big[ \cL_\al \big] \,. \]

\mypar{(D) Variance of the loss (exact).} To begin, let us write $\cD$ for the discrete uniform distribution over $\{0, \pi/2\}^m$. We prove, by induction on the number of parameters $m$, that
\begin{equation}\label{eq:induction_T}
\E_{\T} \left[ \cL_\al^2 \right] = \E_\cD \left[ \cL_\al^2 \right] \coloneqq \left( \frac{1}{2} \right)^m \sum_{\T \in \{0, \pi/2\}^m} \cL_\al^2(\T) \,,
\end{equation}
The base case holds trivially since there are no parameters to integrate over. For the induction step, let $U_m$ be a circuit with $m \geq 1$ parameters, and $\T' = (\T_1, \ldots, \T_m)$. Using the circuit structure from Definition \ref{def:circuit}, we can write
\[ U_m(\T) = U_{m-1}(\T') A_m(\T_m) \coloneqq U_{m-1}(\T') W_m P_m(\T_m) \,, \]
where $W_m$ is a Clifford gate (possibly the identity), and $P_m$ is an arbitrary Pauli string.
Now let $\ga$ be such that $P_\ga = W_m^\dagger P_\al W_m$. If $P_m$ commutes with $P_\ga$, we have $P_m(\T) P_\ga = P_\ga P_m(\T)$ and in particular,
\[ A_m^\dagger(0) P_\al A_m(0) + A_m^\dagger(\pi/2) P_\al A_m(\pi/2) = 2 P_\ga \,. \]
This implies that we can rewrite
\begin{align*}
\E_{\T} \left[ \cL_\al^2 \right] &= \E_{\T'} \left[ \Tr( U_{m-1}^\dagger P_\ga U_{m-1} \rho)^2 \right] \\
&= \frac{1}{2} \sum_{\T_m \in \{0, \pi/2\}} \E_{\T'} \left[ \Tr( U_{m-1}^\dagger A_m^\dagger(\T_m) P_\al A_m(\T_m) U_{m-1} \rho)^2 \right] \,.
\end{align*}
Similarly, if $P_m$ does not commute with $P_\ga$, we have $P_m(\T) P_\ga = P_\ga P_m(-\T)$ and thus
\[ A_m^\dagger(0) P_\al A_m(0) + A_m^\dagger(\pi/2) P_\al A_m(\pi/2) = P_\ga + P_m(-\pi) P_\ga = P_\ga + i P_m P_\ga \,. \]
Using Lemma \ref{lem:trig} to integrate over $\T_m$, we can again rewrite the expectation in the same form:
\begin{align*}
\E_{\T} \left[ \cL_\al^2 \right] &= \E_{\T} \left[ \Tr( U_{m-1}^\dagger P_m(-2\T_m) P_\ga U_{m-1} \rho)^2 \right] \\
&= \E_{\T_m} \left[ \cos^2\T_m \right] \E_{\T'} \left[ \Tr( U_{m-1}^\dagger P_\ga U_{m-1} \rho)^2 \right] + \E_{\T_m} \left[ \sin^2 \T_m \right] \E_{\T'} \left[ \Tr( U_{m-1}^\dagger i P_m P_\ga U_{m-1} \rho)^2 \right] \\
&= \frac{1}{2} \sum_{\T_m \in \{0, \pi/2\}} \E_{\T'} \left[ \Tr( U_{m-1}^\dagger A_m^\dagger(\T_m) P_\al A_m(\T_m) U_{m-1} \rho)^2 \right] \,.
\end{align*}
By the induction hypothesis, since $A_m^\dagger(\T_m)P_\al A_m(\T_m)m$ is a Pauli string for each $\T_m \in \{0, \pi/2\}$, we obtain
\begin{align*}
\E_{\T} \left[ \cL_\al^2 \right] &= \frac{1}{2} \sum_{\T_m \in \{0, \pi/2\}} \frac{1}{2^{m-1}} \sum_{\T' \in \{0, \pi/2\}^{m-1}} \Tr( U_{m-1}^\dagger(\T') A_m^\dagger(\T_m) P_\al A_m(\T_m) U_{m-1}(\T') \rho)^2 \\
&= \left( \frac{1}{2} \right)^m \sum_{\T \in \{0, \pi/2\}^m} \Tr( U_m^\dagger(\T) P_\al U_m(\T) \rho)^2 = \left( \frac{1}{2} \right)^m \sum_{\T \in \{0, \pi/2\}^m} \cL_\al^2(\T, \TT, \TTT) \,.
\end{align*}
This concludes the induction.\\

\mypar{(E) Variance of the loss (bounds).} Recall, from Equation \eqref{eq:circuit_rewrite}, that we can decompose the circuit as $U_K = A_K(\T) B(\TT) C(\TTT)$, with $A_K(\T) = \prod_{k=1}^K W_k P_k(\T_k)$. In particular, for any fixed $\TT, \TTT$, Equation \eqref{eq:induction_T} gives
\begin{equation}\label{eq:induction_T_mod}
\E_{\T} \left[ \cL_\al^2(\T, \TT, \TTT) \right] = \left( \frac{1}{2} \right)^K \sum_{\T \in \{0, \pi/2\}^K} \cL_\al^2(\T, \TT, \TTT) \,,
\end{equation}
where $K = m - 2n$. Now notice that $P_k(\T_k)$ are Clifford gates for all $\T_k \in \{0, \pi/2\}$, implying that
\[ P_{\al(\T)} \coloneqq A_K^\dagger(\T) P_\al A_K(\T) \]
is a Pauli string for all $\T \sim \cD$, whose index we denote by $\al(\T)$. Assuming $\rho = \otimes \rho_i$ is a product mixed state, Equation \eqref{eq:induction_T_mod} now decomposes into the following product:
\begin{align}
\E \left[ \cL_\al^2 \right] 
&= \left( \frac{1}{2} \right)^K \sum_{\T \in \{0, \pi/2\}^K} \ \prod_i \ \E \left[ \cL_\al^i(\T, \TT, \TTT)^2 \right] \label{eq:product} \\
&\coloneqq \left( \frac{1}{2} \right)^K \sum_{\T \in \{0, \pi/2\}^K} \ \prod_i \ \E \left[ \Tr(\sig_{\nu_i}(-\TTT_i) \sig_{\mu_i}(-\TT_i) \sig_{\al_i(\T)} \sig_{\mu_i}(\TT_i) \sig_{\nu_i}(\TTT_i) \rho_i)^2 \right] \nonumber \,.
\end{align}
If $\al_i(\T) = \mu_i$, the $i$th term reduces to
\[ \E \left[ \cL_\al^i(\T, \TT, \TTT)^2 \right] = \E \left[ \Tr(\sig_{\nu_i}(-\TTT_i) \sig_{\al_i(\T)} \sig_{\nu_i}(\TTT_i) \rho_i)^2 \right] \,, \]
and, by assumption that $\al_i(\T) = \mu_i \neq \nu_i$, integrating over $\TTT_i$ using Lemma \ref{lem:trig} yields
\begin{equation}
\E \left[ \cL_\al^i(\T, \TT, \TTT)^2 \right] = \frac{1}{2} \Tr(\sig_{\al_i(\T)} \rho_i)^2 + \frac{1}{2} \Tr(i \sig_{\nu_i} \sig_{\al_i(\T)} \rho_i)^2 \label{eq:variance_first_case} \,.
\end{equation}
In particular, by decomposing $\rho_i$ into the Pauli basis, we can notice that
\begin{equation}
\sum_{j \neq 0} \Tr(\sig_j \rho_i)^2 = 2\Tr(\rho^2_i) - 1 \leq 1 \,, \label{eq:rho_pauli} 
\end{equation}
and hence,
\[ \E \left[ \cL_\al^i(\T, \TT, \TTT)^2 \right] \leq \frac{1}{2} \sum_{j \neq 0} \Tr(\sig_j \rho_i)^2 \leq \frac{1}{2} \,. \]
Otherwise, if $\al_i(\T) \neq 0, \mu_i$, integrating over $\TT_i$ using Lemma \ref{lem:trig} yields
\begin{equation*}
\E \left[ \cL_\al^i(\T, \TT, \TTT)^2 \right] = \frac{1}{2} \E_{\TTT_i} \left[ \Tr(\sig_{\nu_i}(-\TTT_i) \sig_{\al_i(\T)} \sig_{\nu_i}(\TTT_i) \rho_i)^2 \right] + \frac{1}{2} \E_{\TTT_i} \left[ \Tr(\sig_{\nu_i}(-\TTT_i) i \sig_{\mu_i} \sig_{\al_i(\T)} \sig_{\nu_i}(\TTT_i) \rho_i)^2 \right] \,.
\end{equation*}
Now notice that $\al_i(\T) \neq \mu_i \neq \nu_i$ implies $\sig_{\al_i(\T)} = \sig_{\nu_i}$ and $i \sig_{\mu_i} \sig_{\al_i} \neq \sig_{\nu_i}$ or vice-versa. In the former case (wlog), the first term simplifies without integration, while integrating over $\TTT_i$ in the second term yields
\begin{align}
\E \left[ \cL_\al^i(\T, \TT, \TTT)^2 \right]
&= \frac{1}{2} \Tr( \sig_{\al_i(\T)} \rho_i)^2 + \frac{1}{4} \Tr(i \sig_{\mu_i} \sig_{\al_i(\T)} \rho_i)^2 + \frac{1}{4} \Tr(i \sig_{\nu_i} i \sig_{\mu_i} \sig_{\al_i(\T)} \rho_i)^2 \nonumber \\
&= \frac{1}{2} \Tr( \sig_{\al_i(\T)} \rho_i)^2 + \frac{1}{4} \Tr(i \sig_{\mu_i} \sig_{\al_i(\T)} \rho_i)^2 + \frac{1}{4} \Tr(\sig_{\mu_i} \rho_i)^2 \label{eq:variance_second_case} \,. 
\end{align}
Again, the fact that all three Pauli terms are distinct implies
\[ \E \left[ \cL_\al^i(\T, \TT, \TTT)^2 \right] \leq \frac{1}{2} \sum_{j \neq 0} \Tr(\sig_j \rho_i)^2 \leq \frac{1}{2} \,. \]
We thus obtain $\E \left[ \cL_\al^i(\T, \TT, \TTT)^2 \right] \leq \frac{1}{2}$ for all $\al_i(\T) \neq 0$, and taking the product over all qubits yields
\[ \prod_i \E \left[ \cL_\al^i(\T, \TT, \TTT)^2 \right] \leq \left(\frac{1}{2}\right)^\abs{\al(\T)} \]
in the usual $L_0$ pseudo-norm $\abs{\al} \coloneqq \abs{ \{ \al_i \neq 0 \} }$. Now recall that the light-cone $\cone_\al^\bT$ denotes the number of qubits on which $U^\dagger(\bT) P_\al U(\bT)$ acts non-trivially. Since the first two layers $B(\phi), C(\om)$ only contain single-qubit rotations, which cannot alter light-cones, this is equivalent to defining $\cone_\al^\bT = \conealphatheta$ as the number of qubits on which $A_K^\dagger(\T) P_\al A_K(\T)$ acts non-trivially. 
In particular, for all $\T \sim \cD$, we have
\[ P_{\al(\T)} \coloneqq A_K^\dagger(\T) P_\al A_K(\T) \]
and thus $\conealphatheta = \abs{\al(\T)}$. Re-invoking Equation \eqref{eq:product}, we thus conclude
\begin{align}\label{eq:upper_bound}
\E \left[ \cL_\al^2 \right] \leq \left( \frac{1}{2} \right)^K \sum_{\T \in \{0, \pi/2\}^K} \left(\frac{1}{2}\right)^\conealphatheta = \E_\cD \left[ \ \left(\frac{1}{2}\right)^\conealphatheta \ \right] \,.
\end{align}
For the lower bound, let us define the orthogonality between $\rho$ and the first layer of rotations $\otimes_i R_{\sig_{\nu_i}}$ as
\[ \Om(\rho) = \sum_{\substack{\la \in \{0, 1, 2, 3\}^n \\ \forall i \, : \, \la_i \neq 0, \nu_i}} \Tr(P_\la \rho)^2 \,.\]
Now notice that for any $\al(\T) \neq 0$, equations \eqref{eq:variance_first_case} and \eqref{eq:variance_second_case} both yield a lower bound
\[ \E \left[ \cL_\al^i(\T, \TT, \TTT)^2 \right] \geq \frac{1}{4} \Tr(\sig_{\be_i} \rho_i)^2 + \frac{1}{4} \Tr(\sig_{\ga_i} \rho_i)^2 = \frac{1}{4} \sum_{\la_i \neq 0,\nu_i} \Tr(\sig_{\la_i} \rho_i)^2 \,, \]
where $\be_i, \ga_i$ are indices such that $\be_i \neq \ga_i \neq \nu_i \neq 0$.

On the other hand, if $\al_i(\T) = 0$, we decompose $\rho_i = \sum_j c_{ij} \sig_j$ into the Pauli basis and notice that $\Tr(\rho_i) = 1 = 2c_0$ implies
\[ \Tr(\rho_i^2) \geq 2 \Tr(\rho_i^2)-1 = \sum_{\la_i \neq 0} \Tr(\sig_{\la_i} \rho_i)^2 \geq \sum_{\la_i \neq 0,\nu_i} \Tr(\sig_{\la_i} \rho_i)^2 \,. \]
Invoking Equation \eqref{eq:rho_pauli}, we therefore have 
\[ \E \left[ \cL_\al^i(\T, \TT, \TTT)^2 \right] = \Tr(\rho_i)^2 \geq \sum_{\la_i \neq 0,\nu_i} \Tr(\sig_{\la_i} \rho_i)^2 \,. \]
Taking the product over all qubits, we obtain
\[ \E \left[ \cL_\al(\T, \TT, \TTT)^2 \right] \geq \left( \frac{1}{4} \right)^\abs{\al(\T)} \, \prod_{i} \, \sum_{\la_i \neq 0,\nu_i} \Tr(\sig_{\la_i} \rho_i)^2 = \left( \frac{1}{4} \right)^\conealphatheta \Om(\rho) \]
and thus, using Equation \eqref{eq:product} once more,
\[ \E \left[ \cL_\al^2 \right] \geq \left( \frac{1}{2} \right)^K \sum_{\T \in \{0, \pi/2\}^K} \left(\frac{1}{4}\right)^\conealphatheta \Om(\rho) = \Om(\rho) \, \E_\cD \left[ \ \left(\frac{1}{4}\right)^\conealphatheta \ \right] \,. \]
Rewriting our upper and lower bounds in terms of variances, we obtain, for any $\al \neq 0$,
\begin{gather*}
\Om(\rho) \, \E_{\cD} \left[ \  \left( \frac{1}{4} \right)^{\conealphatheta} \ \right] \leq \Var \big[ \cL_\al \big] \leq \E_{\cD} \left[ \ \left( \frac{1}{2} \right)^{\conealphatheta} \ \right] \,.
\end{gather*}
In the special case where $\rho = \ketbra{0}{0}^{\otimes n}$ is the zero initial state and the first rotations are not $Z$-rotations, note that $\rho$ is pure and fully orthogonal to the first layer, so that $\Om(\rho) = 1$.\\

\mypar{(F) Tightness of bounds.} Our concentration bounds can be attained by first setting $K = 0$, where the circuit is no more than two layers of orthogonal single-qubit rotations:
\[ U(\TT, \TTT) = \bigotimes_{i=1}^n \sig_{\mu_i}(\TT_i) \sig_{\nu_i}(\TTT_i) \,. \]
In this case, single-qubit rotations do not alter light-cones, hence $\conealphatheta = \abs{\al}$ for all $\T \sim \cD$ and the bounds reduce to
\begin{gather*}
\Om(\rho) \left( \frac{1}{4} \right)^{\abs{\al}}  \leq \Var \big[ \cL_\al \big] \leq \left( \frac{1}{2} \right)^{\abs{\al}} \,.
\end{gather*}
The upper bound is now reached by pairing this circuit with any Pauli string $P_\al$ such that $\al_i = \mu_i$ for all $\al_i \neq 0$. Similarly, the lower bound is reached whenever $\rho$ is a pure product state and $\al_i \neq \mu_i$ for all $i$.\\

\mypar{(G) Extension to arbitrary mixed states.} We extend our result to arbitrary initial states $\rho$, by decomposing $\rho = \sum_\la d_\la P_\la$ into the Pauli basis and defining a generalized $\al$-orthogonality measure $\Om(\rho, \al)$ as
\[ \Om(\rho, \al) \coloneqq \sum_{\la \in \La(\al)} \Tr\big(P_\la \rho \big)^2 = \sum_{\la \in \La(\al)} 4^n d_\la^2 \,, \]
where
\[ \La(\al) = \{ \la \in \{0, 1, 2, 3\}^n \mid \la_i = 0 \ \ \forall \al_i = 0 \quad \land \quad \la_i \neq 0, \nu_i \ \ \forall \al_i = \mu_i  \quad \land \quad \la_i \neq 0 \ \ \text{otherwise} \} \,. \]
Intuitively, $\Om(\rho, \al)$ measures the `portion' of $\rho$ which is orthogonal to the first layer, on each qubit line where $P_\al$ does not commute with the second layer (namely $\al_i \neq 0, \mu_i$), by summing over the respective coefficients $d_\la^2$.
We generalize our bounds by first proving that for any mixed state $\rho$, and any $\al \neq 0$,
\[ \Om(\rho, \al) \left(\frac{1}{4}\right)^\abs{\al}   \ \leq \  \E_{\TT, \TTT} \left[ \Tr(C^\dagger B^\dagger P_\al B C \rho)^2 \right] \ \leq \ \Om(\rho, \al) \left(\frac{3}{4}\right)^\abs{\al} \,. \]
To begin, we let $r = \abs{ \{ \al_i \neq 0, \mu_i \}}$ be the number of qubit lines on which $P_\al$ does not commute with the second layer, and integrate over corresponding parameters $\TT_i$ using Equation \eqref{eq:induction_T} to obtain
\begin{equation}\label{eq:generalized}
\E_{\TT, \TTT} \left[ \Tr\left(C^\dagger B^\dagger P_{\al} B C \rho\right)^2 \right]
= \left(\frac{1}{2}\right)^r \sum_{\phi \in \{0, \pi/2\}^r} \E_{\TTT} \left[ \Tr\left(C^\dagger P_{\al(\phi)} C \rho\right)^2 \right] \,,
\end{equation}
where $\al(\phi)$ is the index of the Pauli string
\[ P_{\al(\phi)} = \bigotimes_{\al_i = 0, \mu_i} \sig_{\al_i} \bigotimes_{\al_i \neq 0, \mu_i} \sig_{\mu_i}(-\phi_i) \sig_{\al_i} \sig_{\mu_i}(\phi_i) \,. \]
Let us now fix any $\be \coloneqq \al(\phi)$. As above, we let $t = \abs{ \{\be_i \neq 0, \nu_i\} }$ be the number of qubit lines on which $P_\be$ does not commute with the first layer, and define $\be(\TTT)$ as the index of the Pauli string
\[ P_\be = \bigotimes_{\be_i = 0, \nu_i} \sig_{\be_i} \bigotimes_{\be_i \neq 0, \nu_i} \sig_{\nu_i}(-\TTT_i) \sig_{\be_i} \sig_{\nu_i}(\TTT_i) \]
for $\TTT \in \{0, \pi/2\}^t$.
Similarly, we now integrate over parameters $\TTT_i$ using Equation \eqref{eq:induction_T} to obtain
\begin{align*}
\E_{\TTT} \left[ \Tr(C^\dagger P_{\be} C \rho)^2 \right] &= \left( \frac{1}{2} \right)^t \sum_{\TTT \in \{0, \pi/2\}^t} \left( \sum_\la d_\la \Tr( P_\be(\TTT) P_\la) \right)^2 \\
&= \left( \frac{1}{2} \right)^t \sum_{\TTT \in \{0, \pi/2\}^t} \left( \sum_\la 2^n 
d_\la \del(\be(\TTT) = \la) \right)^2 \\
&= \left( \frac{1}{2} \right)^t \sum_{\TTT \in \{0, \pi/2\}^t} \sum_\la 4^n d_\la^2 \del(\be(\TTT) = \la) \,. 
\end{align*}
In order to pick out the non-zero coefficients in this sum over $\TTT$, we notice that $\be_i(0) \neq \be_i(\pi/2) \neq \nu_i \neq 0$ for all $\be_i \neq 0, \nu_i$, and thus define a finer-grained notion of orthogonality
\[ \Ga(\be) = \{ \la \in \{0, 1, 2, 3\}^n \mid \la_i = \be_i \ \ \forall \be_i = 0, \nu_i \quad \land \quad \la_i \neq 0, \nu_i \ \ \text{otherwise} \} \]
to conclude
\begin{align*}
\E_{\TTT} \left[ \Tr(C^\dagger P_{\be} C \rho)^2 \right] = \left( \frac{1}{2} \right)^t \sum_{\la \in \Ga(\be)} 4^n d_\la^2 \,.
\end{align*}
Returning to Equation \eqref{eq:generalized}, and recalling that $r = \abs{\al} - \abs{\al_i = \mu_i}$ and $t = \abs{\al} - \abs{\al(\phi)_i = \nu_i}$, we obtain
\[ \E_{\TT, \TTT} \left[ \Tr\left(C^\dagger B^\dagger P_{\al} B C \rho\right)^2 \right] = \left(\frac{1}{2}\right)^{2\abs{\al}-\abs{\al_i = \mu_i}} \sum_{\phi \in \{0, \pi/2\}^r} 2^{\abs{\al_i(\phi) = \nu_i} } \sum_{\la \in \Ga(\al(\TT))} 4^n d_\la^2 \,. \]
Now, by orthogonality of the first two layers, let $\eta_i$ be the indices such that $0 \neq \eta_i \neq \mu_i \neq \nu_i$ for each $i$. In particular, for any $\al_i \neq 0, \mu_i$, we either have $\al_i = \nu_i$ or $\al_i = \eta_i$, and in turn, $\sig_{\mu_i}(-\phi_i) \sig_{\al_i} \sig_{\mu_i}(\phi_i)$ must take values $\sig_{\nu_i}$ and $\sig_{\eta_i}$ for $\phi_i = 0$ and $\phi_i = \pi/2$ respectively (relabeling parameters if necessary). By construction, we therefore have
\[ \Ga(\al(\TT)) = \{ \la \mid \la_i = 0 \ \ \forall \al_i = 0 \quad \land \quad \la_i \neq 0, \nu_i \ \ \forall \al_i = \mu_i \quad \land \quad \la_i = \nu_i \ \ \forall \phi_i = 0 \quad \land \quad \la_i \neq 0, \nu_i \ \ \forall \phi_i = \pi/2 \} \]
and thus
\[ \bigcup_{\TT \in \{0, \pi/2\}^r} \Ga(\al(\TT)) = \La(\al) \,. \]
For the lower bound, this implies
\begin{align*}
\E_{\TT, \TTT} \left[ \Tr\left(C^\dagger B^\dagger P_{\al} B C \rho\right)^2 \right] &\geq \left(\frac{1}{2}\right)^{2\abs{\al}-\abs{\al_i = \mu_i}} \sum_{\phi \in \{0, \pi/2\}^r} \sum_{\la \in \Ga(\al(\TT))} 4^n d_\la^2 \\
&\geq \left(\frac{1}{2}\right)^{2\abs{\al}-\abs{\al_i = \mu_i}} \sum_{\phi \in \{0, \pi/2\}^r} \sum_{\la \in \cup \Ga(\al(\TT))} 4^n d_\la^2 \\
&\geq \left(\frac{1}{4}\right)^{\abs{\al}} \Om(\al, \rho) \,.
\end{align*}
For the upper bound, we first use $\Ga(\al(\TT)) \subseteq \La(\al)$ and $\abs{\al_i(\phi) = \nu_i} = \abs{\phi_i = 0}$ in order to obtain
\begin{align*}
\E_{\TT, \TTT} \left[ \Tr\left(C^\dagger B^\dagger P_{\al} B C \rho\right)^2 \right] &\leq \left(\frac{1}{2}\right)^{2\abs{\al}-\abs{\al_i = \mu_i}} \sum_{\phi \in \{0, \pi/2\}^r} 2^{\abs{\phi_i = 0}} \sum_{\la \in \La(\al)} 4^n d_\la^2 \\
&= \Om(\rho, \al) \left(\frac{1}{2}\right)^{\abs{\al}+r} \sum_{\phi \in \{0, \pi/2\}^r} 2^{\abs{\phi_i = 0}} \,.
\end{align*}
We now use a standard binomial expansion, along with the fact that $r \leq \abs{\al}$, to conclude
\begin{align*}
\E_{\TT, \TTT} \left[ \Tr\left(C^\dagger B^\dagger P_{\al} B C \rho\right)^2 \right] &\leq \Om(\al, \rho) \left(\frac{1}{2}\right)^{\abs{\al}+r} \sum_{k = 0}^r \binom{r}{k} 2^k \\
&= \Om(\al, \rho) \left(\frac{1}{2}\right)^{\abs{\al}+r} (2+1)^r \\
&\leq \Om(\al, \rho) \left(\frac{3}{4}\right)^{\abs{\al}} \,.
\end{align*}
Returning to the full loss using Equation \eqref{eq:induction_T_mod}, we have
\begin{align*}
\E \left[ \cL_\al^2 \right] = \left(\frac{1}{2}\right)^K \sum_{\T \in \{0, \pi/2\}^K} \E_{\TT, \TTT} \left[ \Tr(C^\dagger B^\dagger P_{\al(\T)} B C \rho)^2 \right] \,,
\end{align*}
allowing us to conclude, writing $\Om_\al^\T = \Om(\rho, \al(\T))$,
\begin{equation*}
\E_\cD \left[ \ \Om_\al^\T \left(\frac{1}{4}\right)^\conealphatheta \ \right] \ \leq \ \Var \big[ \cL_\al \big] \ \leq \ \E_\cD \left[ \ \Om_\al^\T \left(\frac{3}{4}\right)^\conealphatheta \ \right] \,.
\end{equation*}
Again, the bounds can be estimated classically and efficiently, by Monte Carlo sampling $\T \sim \cD$, provided the coefficients $d_\la$ of the initial state $\rho = \sum_\la d_\la P_\la$ are known.
\end{proof}

\begin{mybox}
\begin{appcorollary}{\ref{cor:su2}}
Let $H$ be any mixed observable on $n$ qubits, $\rho$ the zero initial state, and $U(\T)$ an EfficientSU2 circuit of logarithmic depth $O(\log n)$, pairwise entanglement, and rotation layers $(R_Y, R_Z)$. Then the corresponding loss does not suffer from barren plateaus, namely,
\[ \Var_\T \big[ \cL \big] \in \Omega\left( \frac{1}{\poly (n)}\right) \,. \]
\end{appcorollary}
\end{mybox}

\begin{proof}[Proof (Sketch)]
First note that $\rho = \ketbra{0}{0}^n$ is orthogonal to the first layer of $R_Y$ rotations, implying $\Omega(\rho) = 1$. Now let $U(\T) = \prod_{i=1}^d R_i E$ be the EfficientSU2 circuit, where $R_i$ is a product of single-qubit rotations and $E$ is a pairwise entangling layer. Now recall that $\conealphatheta$ is the number of qubits on which $U(\T)^\dagger P_\al U(\T)$ acts non-trivially. Notice that $E$ can at most extend the light-cone of a Pauli matrix $\sig$ by 2 qubits upwards and downwards, hence broadening the light-cone of a $k$-local string $P_\al$ by at most 4 for each of the $k$ non-identity terms. Meanwhile, the single-qubit rotations in $R_i$ cannot alter light-cones, so it follows by induction on the depth $d$ that
\[ \conealphatheta \leq k + 4 k d \]
for any $k$-local $P_\al$ and any $\T$. In particular, applying Theorem \ref{th:bounds}, we obtain
\[ \Var_\T \big[ \cL_\al \big] \ \geq \ \E_{\cD} \left[ \  \left( \frac{1}{4} \right)^{\conealphatheta} \ \right] \geq \left( \frac{1}{4} \right)^{k+4kd} \,. \]
Now by assumption, $H$ contains a $k$-local string $P_\al$ with $k \in O(1)$ and $c_\al \in \Omega(1/\poly(n))$, which implies, for any logarithmic depth $d \in O(\log n)$, that $k+4kd \in O(\log n)$. Invoking the first part of Theorem \ref{th:bounds}, the contribution of this term to the variance is independent, and vanishes at most polynomially vanishing in $n$, namely,
\[ \Var_\T \big[ \cL \big] \ \geq \ c_\al^2 \left( \frac{1}{4} \right)^{k+4kd} \in \ \ \Omega\left( \frac{1}{\poly(n)}\right) \,. \qedhere \]
\end{proof}

\pagebreak

\begin{mybox}
\begin{lemma}\label{lem:trig}
For any random variable $\T$ uniformly distributed over $[-\pi, \pi]$, we have
\[
\E_\T\left[ \sin\T \right] = \E_\T\left[ \cos\T \right] = \E_\T\left[ \sin\T \cos\T \right] dx = 0 \qquad \text{and} \qquad
\E_\T\left[ \sin^2\T \right] = \E_\T\left[ \cos^2\T \right] = \frac{1}{2} \,. \]
\end{lemma}
\end{mybox}

\begin{proof}
For the left-hand-side, we have
\[ \E_\T \left[ \sin\T \right] = \frac{1}{2\pi} \int_{-\pi}^\pi \sin \T d\T = \frac{1}{2\pi} \Big[ - \cos \T \Big]_{-\pi}^\pi = 0 \,. \]
The analogous argument holds for $\cos \T$ and $\sin \T \cos \T = \frac{1}{2} \sin 2\T$. For the right-hand-side, we invoke the standard identity $\cos^2 \T / \sin^2 \T = (1\pm\cos 2\T)/2$ to obtain
\[ \E_\T \left[ \sin^2 \T \right] = \frac{1}{4\pi} \int_{-\pi}^\pi (1-\cos 2\T) d\T = \frac{1}{4\pi} \Big[ \T - \frac{1}{2} \sin 2\T \Big]_{-\pi}^\pi = \frac{1}{2} \,. \]
The analogous argument holds for $\cos^2 \T$.
\end{proof}

\FloatBarrier
\newpage
\section{Light-Cones and Efficient Evaluation}\label{app:efficient_evaluation}

In this appendix, we provide a visual illustration of light-cones and explain how these notions allow us to tighten lower bounds compared with prior work. We then describe how our bounds can be estimated \textit{efficiently and classically}, and numerically demonstrate how they compare favorably with existing bounds for local 2-designs.\\

\mypar{Light-cones.} For a parameterized circuit $U(\T)$ and a Pauli string $P_\al$, recall that the light-cone $\conealphatheta = \cone_\al(\T)$ is the number of qubits on which $U(\T)^\dagger P_\al U(\T)$ acts non-trivially. In Figure \ref{fig:light_cones} below, we visually illustrate the light-cones arising from different values of $\T 
\sim \cD$, for a specific circuit $U(\T)$ and Pauli observable $P_\al = ZII$. In particular, note that single-qubit gates cannot alter light-cones, so it is sufficient to consider only the final parameter $\T_6$. As shown in the figure, this parameter gives rise to light-cones of different sizes, namely, $\cone_\alpha(0) = 1$ and  $\cone_\alpha(\pi/2) = 3$. Theorem \ref{th:bounds} thus implies, for the zero initial state $\rho = \ketbra{0}{0}$, that the variance of the loss is bounded by
\[ 0.13 \approx \frac{1}{2} \left( \frac{1}{4}\right)^{1} + \frac{1}{2} \left( \frac{1}{4}\right)^{3} = \E_\cD \left[ \left( \frac{1}{4}\right)^{\conealphatheta} \right] \leq \Var \big[ \cL_\al \big] \leq \E_\cD \left[ \left(\frac{1}{2}\right)^{\conealphatheta} \right] = \frac{1}{2} \left( \frac{1}{2}\right)^{1} + \frac{1}{2} \left( \frac{1}{2}\right)^{3} \approx 0.31 \,. \]
This approach produces tighter lower bounds than existing work, which only considers the `full' or `causal' light-cone $\fullcone$ in their bounds, namely, the number of qubits on which $U(\T)^\dagger P_\al U(\T)$ acts across \textit{all} values of $\T$. In this example, the full light-cone is $\fullcone = 3$, which would produce a significantly weaker lower bound given by
\[ \left( \frac{1}{4}\right)^{\fullcone} \approx 0.016 \ll 0.13 \,, \]
almost ten times smaller than our own.\\

\begin{figure*}[h]
\centering
\includegraphics[width=\linewidth]{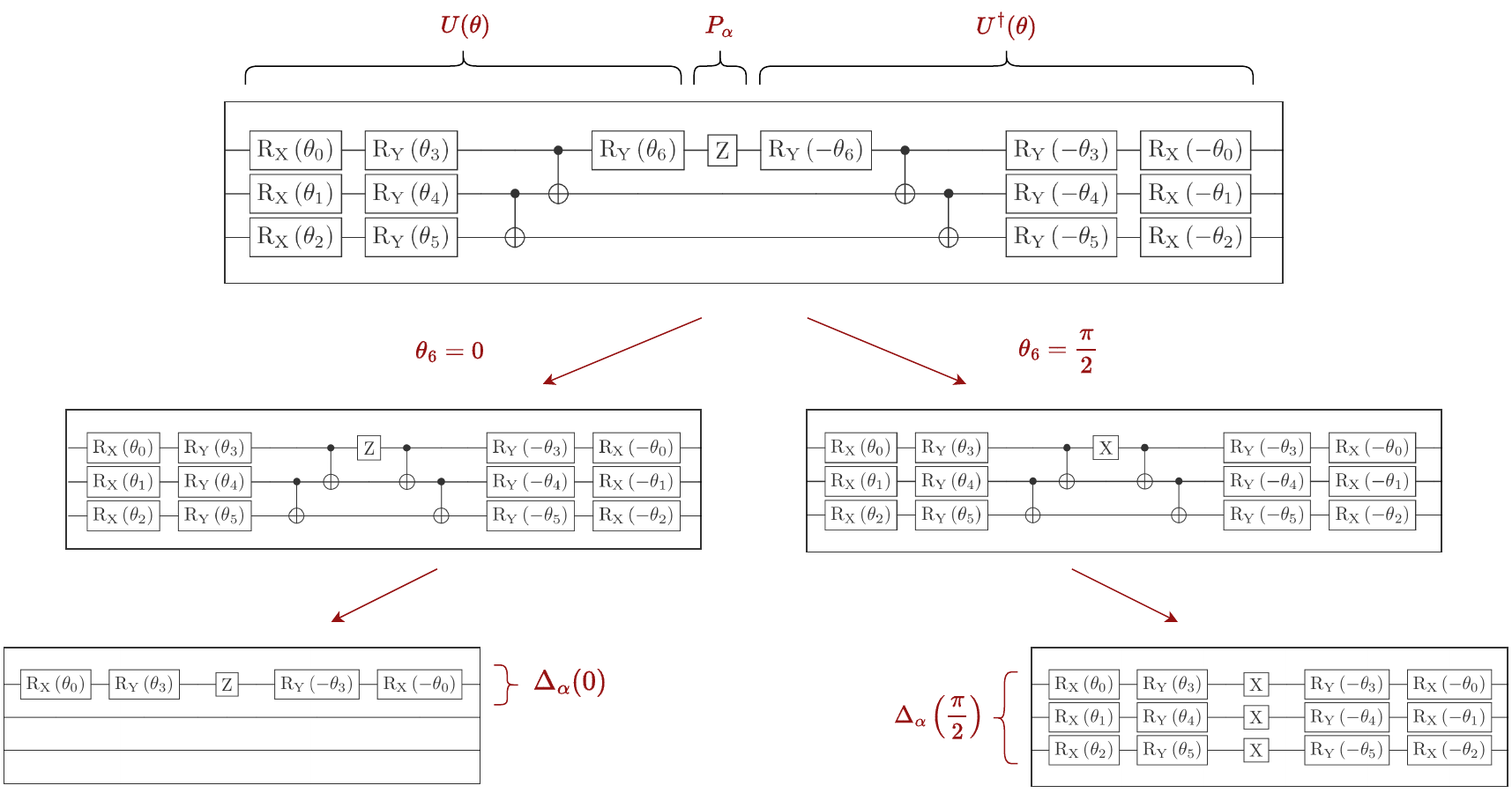}\hspace{5pt}
\caption{Top: diagram representing $U^\dagger P_\al U$, for a fixed circuit $U$ satisfying Definition \ref{def:circuit}, and a Pauli string $P_\al = ZII$. Bottom: corresponding light-cones $\conealpha(\T_6)$, for $\T_6 \in \{0, \pi/2\}$. While $\T_6 = 0$ induces a small light-cone $\conealpha(0) = 1$, because $Z$ gates do not propagate downwards through CNOT gates, $\T_6 = \pi/2$ induces a large light-cone $\conealpha(\pi/2) = 3$ by transforming the $Z$-gate into an $X$-gate, which does propagate.}
\label{fig:light_cones}
\end{figure*}

\newpage

\mypar{Efficient Evaluation.} The key to estimating our bounds efficiently, and classically, is to notice that for any $m$-parameter circuit $U$ from Definition \ref{def:circuit}, $U(\T)$ is a Clifford circuit for any fixed $\T \sim \cD$, recalling that $\cD$ is the discrete uniform distribution over $\{0, \pi/2\}^m$. In particular, for any Pauli string $P_\al$, $U^\dagger(\T) P_\al U(\T)$, is a Pauli string $P_{\al(\T)}$, which can be determined classically with complexity $O(G)$ for a circuit $U$ with $G = \poly(n)$ gates. We can thus efficiently compute the light-cone $\conealphatheta$, namely the number of qubits $i$ with $\al(\T)_i \neq 0$, for any fixed $\T \sim \cD$.
By Monte Carlo sampling, this immediately gives an efficient estimate of the upper and lower bounds
\[ \Om(\rho) \, \E_\cD \left[ \left( \frac{1}{4}\right)^{\conealphatheta} \right] \leq \Var \big[ \cL_\al \big] \leq \E_\cD \left[ \left(\frac{1}{2}\right)^{\conealphatheta} \right] \,, \]
with a convergence speed which is \textbf{independent from the number of qubits}, and a polynomial cost $O(G)$ per sample. Note also that $\Om(\rho)$ need only be computed once, with complexity $O(n)$ for product mixed states.\\

\mypar{Comparison with existing work.} In Figure~\ref{fig:light-cones-bounds-appendix} below, we numerically demonstrate that our bounds are significantly tighter than existing work, both for (a) the EfficientSU2 ansatz and (b) a Cartan ansatz forming an approximate local 2-design \cite{Uvarov_2021BPs}. For further completeness, the experiments are realised for three different pairings of the circuit depth $d$ and the observable locality $k$: (1) linear depth $d = n/2$ and ultralocality $k = 1$, (2) linear depth $d = n/2$ and logarithmic locality $k = \log_2 n$, and (3) logarithmic depth and locality $d = k = \log_2 n$. The observables always act as $Z$ measurements on the first $\log_2 n$ qubits, but analogous experiments with $X$ or $Y$ measurements in different positions demonstrate similar results.\\

Using Theorem~\ref{th:bounds}, the loss variance $\Var_\T \left[ \cL_\al \right] = \E_\cD \left[ \cL^2 \right]$, as well as our upper and lower bounds, are estimated classically with $1000$ samples of $\T \sim \cD$, with the zero initial state $\rho = \ketbra{0}{0}$ satisfying $\Om(\rho)=1$. The shaded areas represent confidence intervals for a confidence level of $0.95$, using the Clopper-Pearson method for the loss variance.
The lines denoted as \textit{lower bound (Uvarov)} and \emph{lower/upper bound (Napp)} represent the bounds derived by \cite{Uvarov_2021BPs} and \cite{napp2022quantifying} respectively. The causal (full) light-cone from \cite{Uvarov_2021BPs} is derived analytically: for any observable $P_\al$ acting non-trivially on the first $k$ qubits, and the EfficientSU2 or Cartan ansatz with $d \geq 1$ layers and $n$ qubits, the full cone is given by
\[ \fullcone = \min\{n, 2\lfloor{k/2}\rfloor + 2d\} \,. \]
Note that we count one layer as containing two sub-layers of pairwise entanglement / Cartan blocks on alternating qubits, as illustrated in Figure \ref{fig:qc_thm_1}, so that the total depth of the circuit is $2d$ in the language of \cite{napp2022quantifying, Uvarov_2021BPs}. As visible in the figure, our bounds are significantly tighter than existing work.\\

It is worth remarking that the bounds of \cite{Uvarov_2021BPs} only apply for local 2-designs, which is not the case for the EfficientSU2 ansatz, and only approximately true for the Cartan ansatz. Similarly, the bounds of \cite{napp2022quantifying} only apply to circuits whose entangling gates are chosen randomly. Finally, the lower bound of \cite{Cerezo_2021_costfunct} only applies to sums of at most 2-local observables, for circuits with local blocks forming 2-designs, hence only applying to our experiments in the ultralocal case ($k=1$), see Figure~\ref{fig:light-cones-bounds-appendix}\textcolor{red!80!black}{(1)}. Similarly, their upper bound only applies to global observables which act non-trivially on every even or odd pair of qubits, which does not apply to any of our experiments, since our only global observable, in Figure \ref{fig:light-cones-bounds-appendix}\textcolor{red!80!black}{(2)}, acts exclusively on the first $n/2$ qubits. Indeed, their bound contradicted the true loss variance in our experiment, so we could not include it in the figure. The corresponding numerical results should therefore be interpreted with caution, but are all the more indicative of the broader applicability of our bounds, as well as their tightness.\\

It is also noteworthy that one would generally expect the variance of the loss, as well as our bounds, to decrease monotonically in the number of qubits. This is the case in most of our experiments, but not for the EfficientSU2 ansatz in the $d=k=\log_2 n$ case; see Figure~\ref{fig:light-cones-bounds-appendix}\textcolor{red!80!black}{(3a)}. This observation was reproduced with a larger number of shots, and can indeed happen if the ansatz structure induces the Pauli observable to propagate through entangling gates in such a way that $(1/4)^\conealphatheta$ becomes \textit{larger} (on average) as the depth increases, not smaller -- even if the average light-cone itself may increase. This motivates further work into designing ansätze in such a way as to maximize the lower bound $(1/4)^\conealphatheta$, all the while retaining high circuit expressivity.

\begin{figure*}[t]
\centering
\begin{subfigure}[c]{0.49\linewidth}
\subcaption{EfficientSU2}
\includegraphics[width=\linewidth]{figs/light-cones/EfficientSU2/1/loss_bounds.pdf} 
\end{subfigure} \hfill
\begin{subfigure}[c]{0.49\linewidth}
\subcaption{Approximate local 2-design}
\includegraphics[width=\linewidth]{figs/light-cones/Cartan/1/loss_bounds.pdf} 
\end{subfigure} \hfill
\begin{subfigure}[c]{0.001\linewidth}
\vspace*{-20pt}
\rotatebox{270}{(1) $d = n/2, \, k = 1$}
\end{subfigure} \hfill

\begin{subfigure}[c]{0.49\linewidth}
\includegraphics[width=\linewidth]{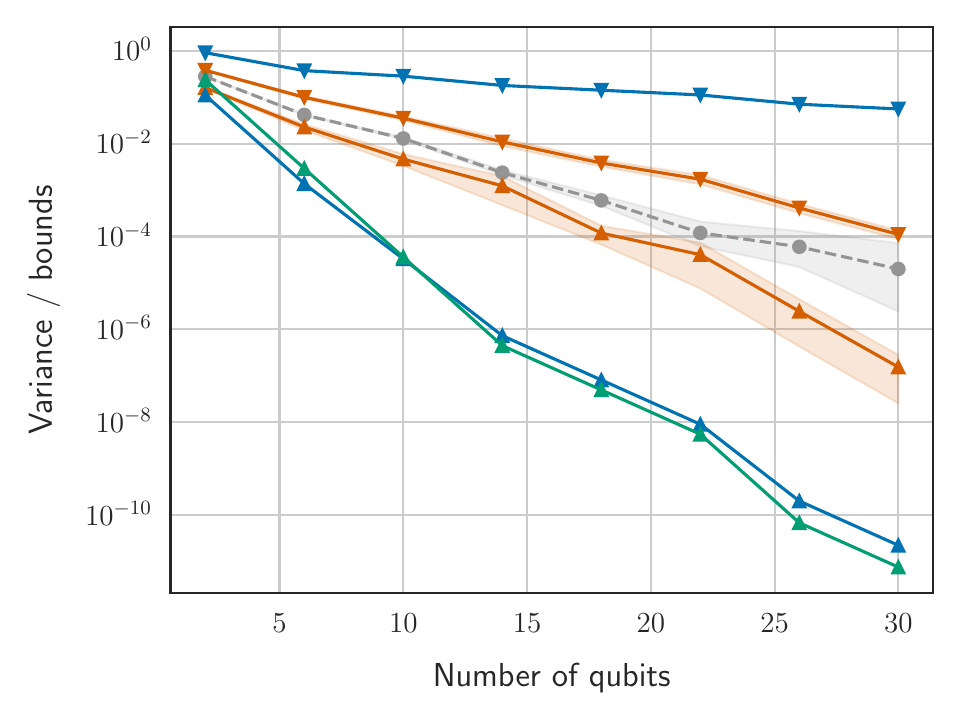} 
\end{subfigure} \hfill
\begin{subfigure}[c]{0.49\linewidth}
\includegraphics[width=\linewidth]{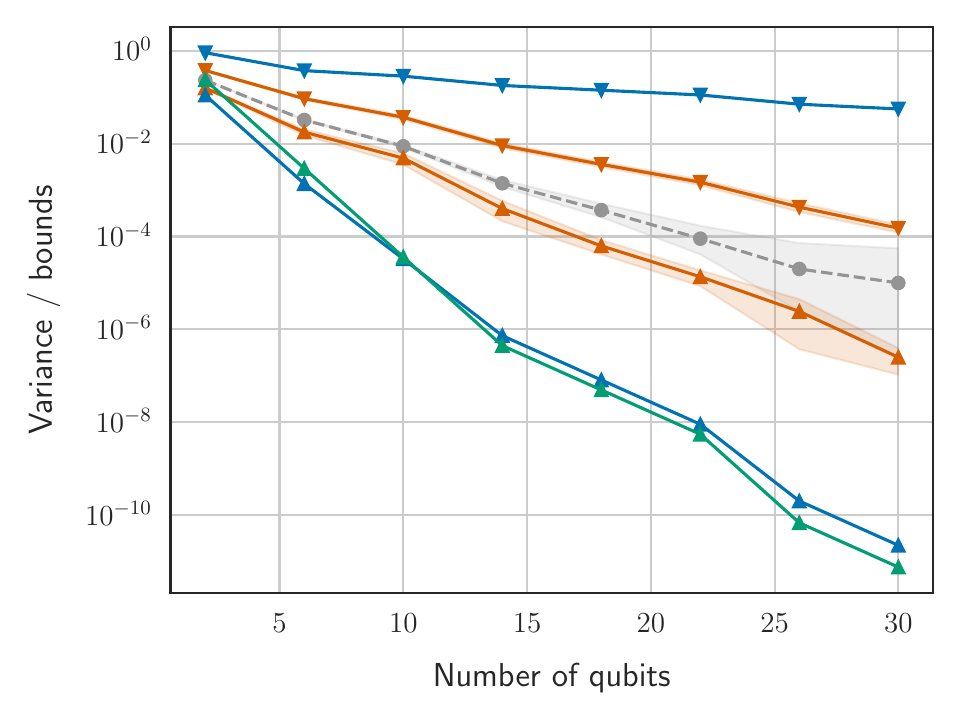} 
\end{subfigure} \hfill
\begin{subfigure}[c]{0.001\linewidth}
\vspace*{-30pt}
\rotatebox{270}{(2) $d = \log_2 n, \, k = n/2$}
\end{subfigure} \hfill

\begin{subfigure}[c]{0.49\linewidth}
\includegraphics[width=\linewidth]{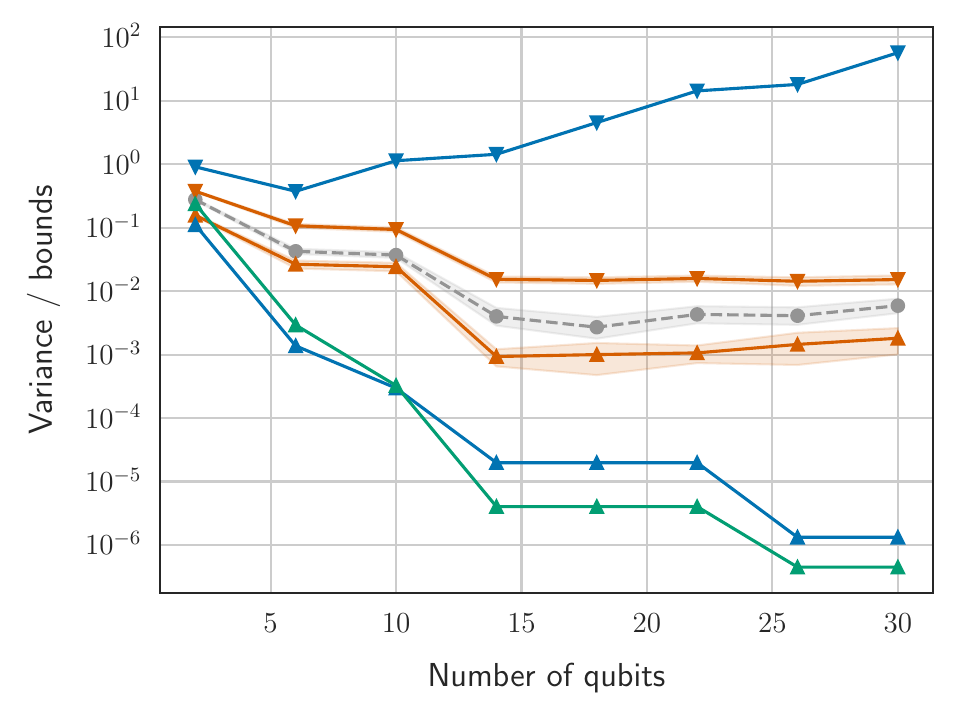} 
\end{subfigure} \hfill
\begin{subfigure}[c]{0.49\linewidth}
\includegraphics[width=\linewidth]{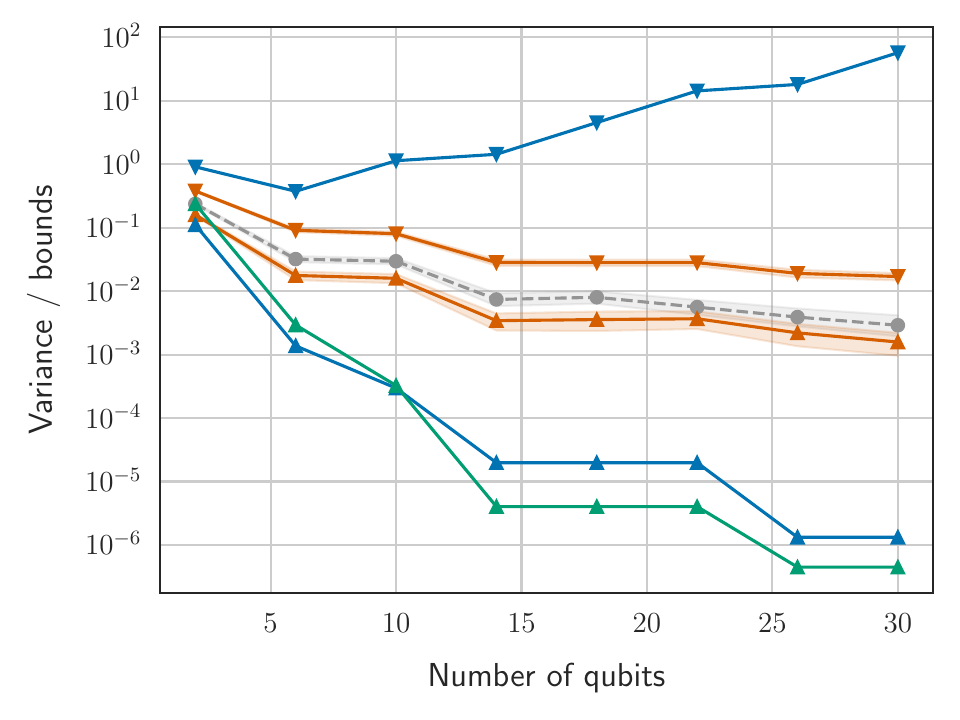} 
\end{subfigure} \hfill
\begin{subfigure}[c]{0.001\linewidth}
\vspace*{-20pt}
\rotatebox{270}{(3) $d = k = \log_2 n$}
\end{subfigure} \hfill
\caption{Variance of the loss function, and corresponding bounds from Theorem \ref{th:bounds}, compared with existing work \cite{napp2022quantifying, Uvarov_2021BPs}, for two ansatz choices (columns a-b) and various pairings (rows 1-3) of circuit depth $d$ and observable locality $k$. (a): EfficientSU2 ansatz. (b): Cartan ansatz forming an approximate local 2-design \cite{Uvarov_2021BPs}. The loss variance is estimated with $10000$ samples of $\T \sim \cD$. Our bounds are also estimated using $10000$ samples for the global observable (1), but only $1000$ samples for rows (2) \& (3), which displayed smaller estimator variance.}
\label{fig:light-cones-bounds-appendix}
\end{figure*}

\FloatBarrier
\mypar{Variance bounds for estimators.} For completeness, we end this appendix with an analysis of our estimators for the lower bound, as well as the variance of the loss itself. First, we look at the variance when estimating $g \coloneqq \E_\cD \big[ \cL^2_\al \big]$ via Monte Carlo sampling. As previously noted, since $U(\T)$ is a Clifford circuit for $\T \sim \cD$, the variable $\cL^2_\al$ takes values in $\{0, 1\}$ over $\cD$. In other words, estimating $g$ corresponds to sampling from a Bernoulli variable $Y$ with success probability $g$. The variance immediately follows as $\Var_\cD \left[ \cL^2_\al \right] = g - g^2$.\\

Second, let us consider the lower bound $h \coloneqq \E_{\cD} [ \ \left( 1/4 \right)^{\conealphatheta} \ ] \leq g$. Since the light-cone $\conealphatheta$ takes values in $\{0, \ldots, n\}$ for $n$ qubits, we approximate this by a Binomial distribution $X$ with parameters $n, p$. Now the Moment Generating Function (MGF) of a (scaled) Binomial random variable $c X$, for a scaling factor $c \in \mathbb{R}$, is given by \cite{Bulmer_1979_statistics}
\[ \E_{\cD} \left[ \  e^{c X t} \ \right] = (1 - p + p e^{c t})^n \,. \]
Thus, setting $t = 1$ and $c=\ln(1/4)$ gives us $h = (1 - 3/4 p)^n$.
Similarly, the variance can be written via the MGF as
\[
\Var\left[ (1/4)^\conealphatheta \right] = (1 - 15/16 p)^n - h^{2} \,.
\]
For small $x$, we can approximate $e^x \approx 1 + x$ to write
\[ 1 - 15/16 x \approx e^{-15/16 x} = e^{(5/4)(-3/4 x)} \approx (1 - 3/4 x)^{5/4} \,. \] Motivated by this relationship, one easily finds that for all $p \in [0, 1]$ and $n \geq 1$, we have
\[
(1 - 15/16 p)^n \leq (1 - 3/4 p)^{5n/4} = h^{5/4} \,,
\]
which implies $\Var[ \, (1/4)^\conealphatheta \, ] \leq h^{5/4}  - h^{2}$. This quantity is monotonically increasing for all $0 \leq h \leq \frac{5^{4/3}}{16} \approx 0.53$, hence
\[
\Var\left[ (1/4)^\conealphatheta \right] \leq g^{5/4}  - g^{2}
\]
for all $g \leq 0.53$. Thus, the lower bound has a variance of $O(g^{5/4})$ while the loss variance has a variance of $O(g)$. For small $g$, this can lead to orders of magnitudes of reduction in the number of samples required to achieve a certain target accuracy. Analogously, the variance of the upper bound can be bounded by $\Var[ \, (1/2)^\conealphatheta \, ] \leq g^{3/2}  - g^{2}$. However, the variance of the upper bound is not maximized when the expectation value equals $g$, so this scaling is only illustrative. Fig.~\ref{fig:variance_bounds} below presents an empirical study to confirm these results.

\begin{figure}[h]
\centering
\includegraphics[width=0.55\linewidth]{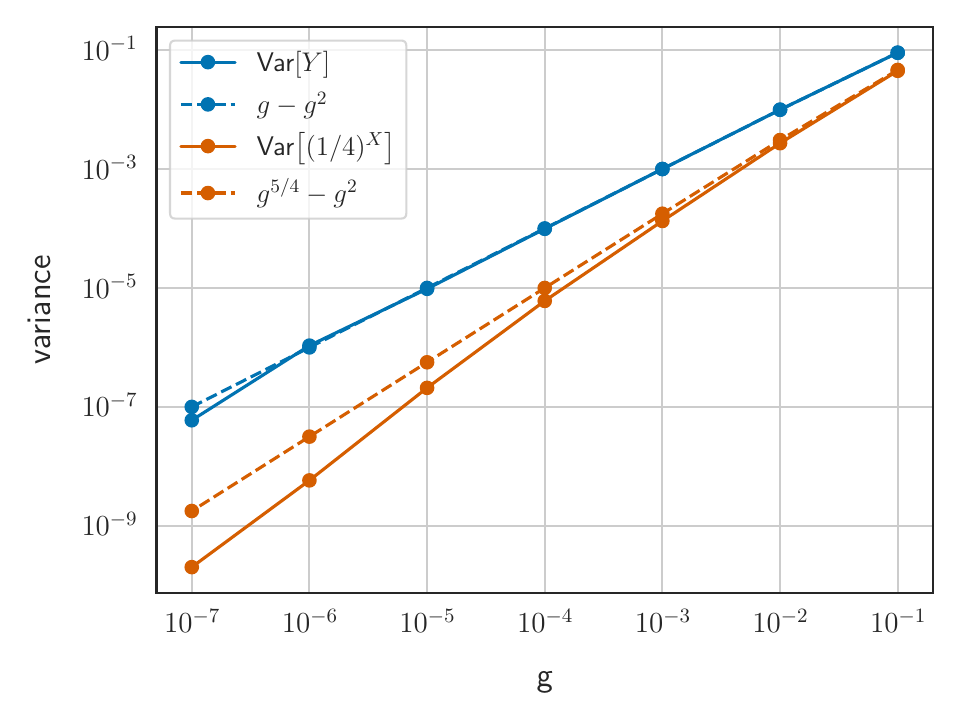} 
\caption{Solid lines: variances of random variables $Y$ and $(1/4)^X$ (solid lines), where $Y$ is Bernoulli with success probability $g$, and $X$ is binomial with parameters $n = 30$ and $p_g = 4/3(1-g^{1/n})$, defined such that $\E\left[(1/4)^X\right] = g$. Dashed lines: analytic bounds derive above, involved in estimating the loss variance and its lower bound in Theorem~\ref{th:bounds}. Every data point, corresponding to $g = 10^{-1}, \ldots, 10^{-7}$, has been estimated using $10^8$ samples.}
\label{fig:variance_bounds}
\end{figure}

\begin{figure*}[t]
\centering
\begin{subfigure}[c]{0.49\linewidth}
\subcaption{EfficientSU2}
\includegraphics[width=\linewidth]{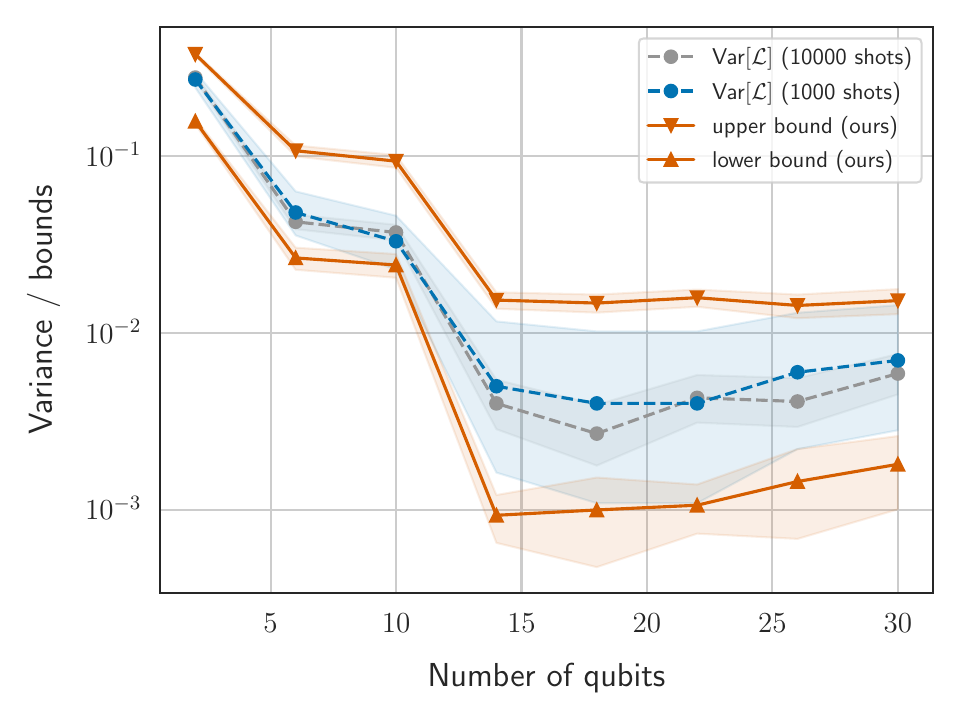} 
\end{subfigure} \hfill
\begin{subfigure}[c]{0.49\linewidth}
\subcaption{Approximate local 2-design}
\includegraphics[width=\linewidth]{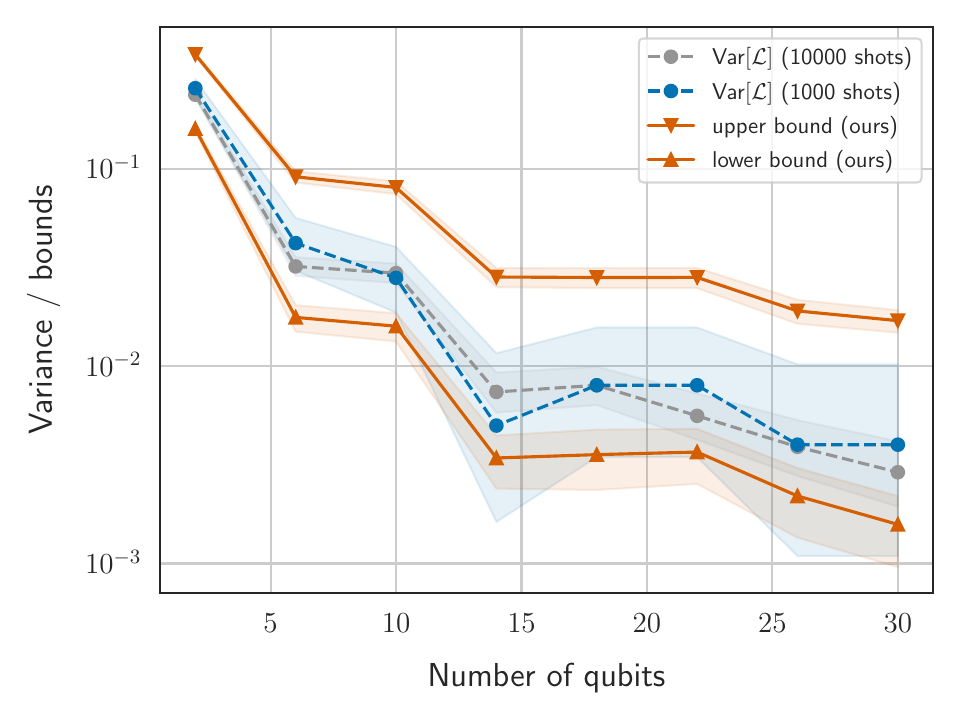} 
\end{subfigure} \hfill
\caption{Variance of the loss function, and corresponding bounds from Theorem \ref{th:bounds}, for two ansatz choices of logarithmic depth and an observable of logarithmic locality. (a): EfficientSU2 ansatz. (b): Cartan ansatz forming an approximate local 2-design \cite{Uvarov_2021BPs}. The bounds are estimated with $1000$ samples of $\T \sim \cD$, while the loss variance is estimated with $1000$ and $10000$ samples of $\T \sim \cD$ respectively.}
\label{fig:variance_variance_loss}
\end{figure*}

To visualize the smaller variance of our lower bound compared with that of the loss variance, Fig.~\ref{fig:variance_variance_loss} above plots the bounds and variances with 1000 samples of $\T \sim \cD$, as well as ten times more (10000) for the loss variance. Notably, as predicted, the loss variance estimation has a significantly larger confidence interval than the bounds for the same number of samples, while reaching a similar confidence interval only for ten times as many samples.

\newpage
\FloatBarrier
\section{Justification of Circuit Assumptions}\label{app:justification}

Here, we provide a justification of the circuit assumptions from Definition \ref{def:circuit}, by providing explicit examples that violate Theorem \ref{th:bounds} whenever any assumption is loosened -- which may be informative in designing circuits skillfully.\newline

\mypar{Orthogonal layers.} As described in Section \ref{sec:general-theory}, the absence of two initial orthogonal layers can induce uniformly zero gradients. A trivial example is to take a circuit consisting only of $R_Y$ rotation gates, and any $Y$-observable, which thus commutes with the circuit entirely. While seemingly artificial, this also applies to the commonly used class of RealAmplitudes circuits \cite{realamp}, which can produce a zero loss when paired with a 1-local $Y$-observable -- although our results can be extended to this class of circuits, provided the corresponding observable contains no $Y$-terms. On the other hand, it is useful to notice that even a circuit which \textit{does have} two orthogonal layers of rotations, but which are \textit{not adjacent}, can also induce a uniformly zero loss. An example is provided in Figure \ref{fig:adjacent} below.

\begin{figure*}[h]
\centering
\includegraphics[height=2.1cm]{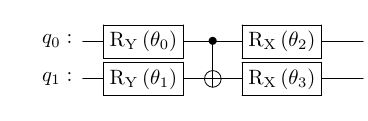}
\caption{Circuit $U$ with orthogonal layers that are not adjacent, inducing a zero loss for $H = XZ$ and $\rho = \ketbra{0}{0}$.}
\label{fig:adjacent}
\end{figure*}

Indeed, by virtue of $Z / X$ gates propagating up/down along CNOT gates, we have
\begin{align*}
U^\dagger H U
&= U^\dagger  \Big( X \otimes Z \Big) \Big( R_X(\T_2) \otimes R_X(\T_3) \Big) \text{CNOT} \Big( R_Y(\T_0) \otimes R_Y(\T_1) \Big) \\
&= U^\dagger  \Big( R_X(\T_2) \otimes R_X(-\T_3) \Big) \Big( X \otimes Z \Big) \text{CNOT} \Big( R_Y(\T_0) \otimes R_Y(\T_1) \Big) \\
&= U^\dagger  \Big( R_X(\T_2) \otimes R_X(-\T_3) \Big) \text{CNOT} \Big( ZX \otimes ZX \Big)  \Big( R_Y(\T_0) \otimes R_Y(\T_1) \Big) \\
&= - \Big( R_Y(-\T_0) \otimes R_Y(-\T_1) \Big) \text{CNOT} \Big( I \otimes R_X(-2\T_3) \Big)  \text{CNOT} \Big( Y \otimes Y \Big) \Big( R_Y(\T_0) \otimes R_Y(\T_1) \Big) \\
&= - \Big( R_Y(-\T_0) \otimes R_Y(-\T_1) \Big)\Big( I \otimes R_X(-2\T_3) \Big) \Big( Y \otimes Y \Big) \Big( R_Y(\T_0) \otimes R_Y(\T_1) \Big) \\
&= - Y \otimes R_Y(-\T_1)  R_X(-2\T_3) Y R_Y(\T_1) \,.
\end{align*}
In particular, $H$ `essentially' commutes with $U$ on the first qubit line, in that $U^\dagger H U$ is a tensor product which is independent from $\T$ on the first component. For the initial state $\rho = \ketbra{0}{0}$, we thus obtain a uniformly zero loss:
\[ \cL(\T) = \Tr(U^\dagger H U \rho) = - \bra{0} Y \ket{0} \times \bra{0} R_Y(-\T_1)  R_X(-2\T_3) Y R_Y(\T_1) \ket{0} = 0 \,. \]

\begin{figure*}[h]
\centering
\begin{subfigure}[b]{0.4\linewidth}
\includegraphics{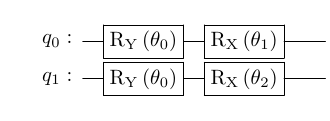}
\subcaption{Dependent parameters.}
\label{fig:dependent}
\end{subfigure}
\begin{subfigure}[b]{0.5\linewidth}
\includegraphics{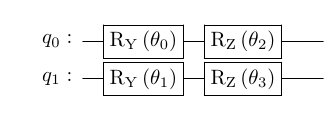}
\subcaption{Parameters initialized uniformly in $[-a\pi, a\pi]$.}
\label{fig:uniform}
\end{subfigure}
\caption{(a) Circuit with dependent parameters, inducing a zero loss for $H = XI - XI$. (b) Circuit inducing a concentrated loss satisfying $\Var [\cL] \leq 5/a^4$, for parameters uniformly distributed in $[-a\pi, a\pi]$ and $H = ZI$.}
\end{figure*}

\mypar{Independent parameters.} If parameters are not independent, different Pauli terms may contribute in opposite directions and lead to uniformly zero gradients. An example is provided in Figure \ref{fig:dependent}.

\mypar{Uniform initialization over $[-\pi, \pi]$.} If parameters are not initialized uniformly over $[-\pi, \pi]$, concentration bounds may still be obtained, but may be significantly poorer. For instance, taking the circuit in Figure \ref{fig:uniform} with the 1-local observable $ZI$ and a uniform distribution over $[-a\pi, a\pi]$ will induce a loss satisfying $\Var [\cL] \leq 5/a^4$. In this work, we have focused on uniform initialization over $[-\pi, \pi]$, but the topic of initialization has been studied extensively and, as demonstrated by \cite{wang2023uniform}, choosing $a$ appropriately can significantly enhance trainability.\\

\begin{figure*}[h]
\centering
\begin{subfigure}[b]{0.4\linewidth}
\includegraphics{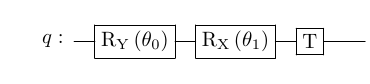}
\vspace{10pt}
\subcaption{Non-Clifford gate.}
\label{fig:clifford}
\end{subfigure}
\begin{subfigure}[b]{0.5\linewidth}
\includegraphics{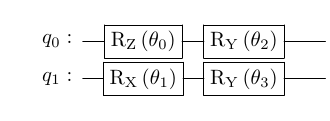}
\subcaption{Initial $Z$-rotation.}
\label{fig:z-rotation}
\end{subfigure}
\caption{(a) Circuit with a non-Clifford gate, where $\E\left[ \cL_\al \cL_\be \right] = \frac{1}{8} \neq 0$ for $P_\al = X$ and $P_\be = Y$. (b) Circuit with an initial $Z$-rotation, inducing a zero loss for the zero initial state and $H = YX$ (in fact, for any $H = Y\sig$).}
\end{figure*}

\mypar{Clifford gates.} If $W_k$ are not Clifford gates, Pauli terms do not necessarily make uncorrelated contributions to the gradient, violating Theorem \ref{th:bounds}. We provide a minimal, single-qubit example in Figure \ref{fig:clifford}, where the covariance fails to vanish due to the presence of a $T$-gate. In particular, the variance no longer decomposes as $\Var\left[ \cL \right] = \sum_\al c_\al^2 \Var\left[ \cL_\al \right]$. 

\mypar{$Z$-rotations and the zero state.} As mentioned below Definition \ref{def:orthogonality}, gradients may vanish if $\rho$ is aligned with an eigenvector of the first-layer rotations. We provide a minimal, single-qubit example in Figure \ref{fig:z-rotation} above, where the zero state commutes with the first $R_Z$ rotation and the observable $H = YX$ commutes with the second $R_Y$ rotation on the first qubit line, producing a uniformly zero loss. Note that neither the initial state nor the observable commutes with rotations on the second qubit line, but the loss is nonetheless uniformly zero. In particular, this motivates our definition of the orthogonality $\Om(\rho)$ as a \textit{product} over each qubit line, and accounting for this allows us to derive a lower bound for all states and initial rotations.

\newpage
\FloatBarrier
\section{Equivalence Between $k$-Degree Polynomials and $k$-Local Diagonal Observables}\label{app:equivalence}

In this appendix, we provide a bijection between binary polynomials $f : \{0, 1\}^n \to \R$ of degree $k$, and diagonal Hermitian observables $H$ whose Pauli terms are at most (algebraically) $k$-local. This implies that generic polynomial binary optimization may include arbitrary global terms -- but Theorem \ref{th:bounds} guarantees that a fixed polynomial may \textit{not} induce barren plateaus, provided it includes a monomial of small degree!\\

\mypar{Context.} The goal of unconstrained black-box binary optimization \cite{Zoufal2023VariationalQuantumBlackBox, IzawaContinuousBlackBox22} is to minimize a function $f: \{0, 1\}^n \rightarrow \mathbb{R}$ which we can evaluate, but whose closed-form expression is unknown. Importantly, this problem can be formulated as a VQA, with a diagonal observable given by $H = \sum_x f\left(x\right) \ket{x}\bra{x}$, where the sum ranges over bitstrings $x$ of length $n$. To decompose this into the Pauli basis, write $Z_\al \coloneqq \bigotimes Z^{\al_i}$ for any $\al \in \{0, 1\}^n$, with $Z^0 = I$ and $Z^1 = Z$. Then,
\begin{equation}\label{eq:weights}
H = \sum_\al c_\al Z_\al \coloneqq \sum_\al \left( \frac{1}{2^n} \sum_x {(-1)}^{\al \cdot x} f(x) \right) Z_\al \,. 
\end{equation}

Since each coefficient $c_\al$ comes with an exponentially small pre-factor $1/2^n$, it is not obvious whether this observable is `mainly' local or global. The prevailing approach has been to check this empirically, as illustrated by \cite{Zoufal2023VariationalQuantumBlackBox}, who show that a number of application-specific observables can be fitted to $2$-local models with high fidelity.\\

However, Theorem \ref{th:bounds} implies that whether or not the observable is `close' to a local one is, in fact, beside the point: all that is necessary to exclude barren plateaus is for \textit{some} local coefficient $c_\al$ to be non-vanishing, \textit{irrespective} of global terms. Moreover, these coefficients can easily (and simultaneously) be estimated by Monte Carlo sampling of the black box $f(x)$, whose convergence rate for $N$ samples is $O(\sqrt{N})$. Crucially, this is \textit{independent} from the number of qubits $n$. If at least one local coefficient is found to be non-vanishing, Corollary \ref{cor:su2} readily guarantees initial trainability for typical shallow circuits. If not, this knowledge could significantly help us design the ansatz. For instance, if a global coefficient is found to be non-vanishing, one may add an appropriate entangling layer, acting as a basis transformation, in order to transform a global term to a local one.\\

A special case of this setting is when $f$ is not a black box but an explicit polynomial $f(x) = \sum_{I \in \cI} c_I x_I$, where $\cI$ is a set of multi-indices, $c_I \in \R$, and $x_I = \prod_{i \in I} x_i$. A well-known instance is quadratic unconstrained binary optimization (QUBO) \cite{kochenberger2014qubo}, where $f$ is required to be \textit{quadratic}. In this case, the corresponding VQA formulation is given by a 2-local observable, which will not induce barren plateaus. In the following proposition, we generalize this correspondence by producing a bijective mapping between polynomials of degree $k$ and $k$-local observables, implying that generic polynomial optimization may include arbitrary global terms. Thankfully, Theorem \ref{th:bounds} guarantees that such observables will \textit{not} induce barren plateaus provided they include a non-vanishing local term, which broadens the spectrum of applications from QUBO to arbitrary polynomial binary optimization.\\

\begin{mybox}
\begin{proposition}\label{prop:equivalence_poly_obs}
For fixed $k, n \in \N$, let $\cF_k$ be the set of polynomials $f : \{0, 1\}^n \to \R$ of degree $k$, and let $\cH_k$ be the set of diagonal Hermitian matrices $H \in M_{2^n}(\C)$ whose Pauli decomposition $H = \sum_\al c_\al Z_\al$ satisfies $\abs{\al} \leq k$ for all $c_\al \neq 0$. Then $\cF_k$ and $\cH_k$ are isomorphic. More specifically, the following map $T : \cF_k \to \cH_k$ is bijective:
\[ T(f) = \sum_{x \in \{0, 1\}^n} f(x) \ket{x}\bra{x} \,. \]
\end{proposition}
\end{mybox}

\begin{proof}
First notice that both sets are finite-dimensional vector spaces over $\R$, since the set of \textit{diagonal} Hermitian matrices is in fact real. Writing $x_\al = \prod_i x^{\al_i}$ as the monomial corresponding to any $\al \in \{0, 1\}^n$, a basis for $\cF_k$ is the set of monomials $\{x_\al\}_{\abs{\al} \leq k}$. Similarly, a basis for $\cH_k$ is the set of Pauli strings $\{ P_\al \}_{\abs{\al} \leq k}$. It follows that $\cH_k$ and $\cF_k$ are isomorphic, since finite-dimensional vector spaces over $\R$ of equal dimension are unique up to isomorphism.\\

To prove that $T$ is bijective, the only non-trivial step is to show that it is well-defined, i.e., $T(f) \in \cH_k$ for all $f \in \cF_k$. Since $T$ is linear, it suffices to show this holds for any monomial $f(x) = x_\al$ with $\abs{\al} \leq k$. Using the decomposition of projectors into the Pauli basis from Equation \eqref{eq:weights}, we have
\[ T(x_\al) = \sum_\be c_\be P_\be \coloneqq \sum_\be \left( \frac{1}{2^n} \sum_x (-1)^{\be \cdot x} x_\al \right) P_\be \,. \]
We must now prove that $c_\be = 0$ for any $\abs{\be} > k$. For any such $\be$, since $\abs{\al} \leq k$, we must have $\be_i = 1$ and $\al_i = 0$ for some index $i$. Writing $x', \al', \be'$ for the bitstrings corresponding to $x, \al, \be$ with the $i$th term omitted, we find
\[ c_\be = \frac{1}{2^n} \sum_{x'} (-1)^{\be' \cdot x'} \sum_{x_i} (-1)^{x_i} x_\al = \frac{1}{2^n} \sum_{x'} (-1)^{\be' \cdot x'} x'_{\al'} \sum_{x_i} (-1)^{x_i} = 0 \,. \]
We therefore conclude $T(x_\al) \in \cH_k$. Now that the mapping is well-defined, it remains to show bijectivity. But since $\cF_k, \cH_k$ have already been shown to be isomorphic, this follows from injectivity. The latter trivially holds, since $T(f) = 0$ implies $f(x) = 0$ for all $x$, by virtue of $\ket{x}\bra{x}$ being a basis for $\cH_n$, and hence $f = 0$.
\end{proof}    

\vspace{5pt}\mypar{Remark.} Any binary function $f : \{0, 1\}^n \to \R$ can be written as a polynomial of degree $n$, by inductively extending the following one-variable identity $f(x) = f(0)(1-x) + f(1)x$. It immediately follows that Proposition \ref{prop:equivalence_poly_obs} extends to all binary functions which can be written as a sum of $k$-\textit{variable} functions, rather than $k$-\textit{degree} monomials. This may be particularly relevant for \textit{black-box} binary optimization, where $f$ is arbitrary.

\newpage
\section{Proof of Theorem \ref{th:qgan}}\label{app:qgan_proof}

\begin{mybox}
\begin{apptheorem}{\ref{th:qgan}}
Consider any discriminator $\cD_\phi \in \cN$, and the corresponding weights $c_\al(\phi)$ defined by Equation \eqref{eq:qgan_coeffs}. Then, for any 1-local weight (namely $\abs{\al} = 1$), we have
\begin{align*}
\E_\phi \Big[ c_\al(\phi)^2 \Big] &\geq \frac{\sig_{L+1}^2}{16} \ \prod_{l=1}^{L} \frac{m_l \sig_l^2(1+\gamma_l)^2}{4} \,.
\end{align*}
In particular, initialising parameters such that $m_l \sig^2_l \geq 4$ for each $l$, the bound reduces to $\E_\phi \big[ c_\al(\phi)^2 \big] \geq \sig^2_{L+1}/16$, which is \textbf{constant} both in the number of qubits $n$ and the discriminator depth $L$.
\end{apptheorem}
\end{mybox}

\begin{proof} Consider a fixed discriminator $D_\phi \in \cN$, and denote its leaky-ReLU activation function by $R(x) = \max(x, \ga x)$ (with parameter $\ga$), as well as its weight / bias matrices by $A^l$ / $B^l$, for each layer $l$, together forming its parameter set $\phi$. We inductively define the output of the $l$th layer as $D^0(x) = x$ for the input,
\[ D^{l}(x) = R \left( A^l D^{l-1}(x) + B^l \right) \]
for the hidden layers $l : 1 \to L$, and 
\[ D^{L+1}(x) = A^{L+1} D^{L}(x) + B^{L+1} \]
for the output (no leaky ReLU). A discriminator with $L$ hidden layers can thus be expressed as
\[ D_\phi(x) = F \left( D^{L+1}(x) \right) \,, \]
where the output activation $F : \R \to \R$ is given by $F(x) = \log(\sig(x))$ for min-max GANs and $F(x) = x$ for Wasserstein GANs. Finally, we write $m_0, \ldots, m_{L+1}$ for the width of each layer, so that $A^l \in \R^{m_l \times m_{l-1}}$ and $B^l \in \R^{m_l}$ for each $l$. Now recall the definition of $c_\al$ from Equation \eqref{eq:weights}:
\[ c_\al = \frac{1}{2^n} \sum_{x \in \{0, 1\}^n} {(-1)}^{\al \cdot x} D_\phi(x) \,. \]
Our goal is to provide a lower bound for
\[ \E_\phi \Big[ c_\al(\phi)^2 \Big] = \E_\phi \left[ \frac{1}{4^n} \sumxy D_\phi(x) \cdot D_\phi(y) \right] \,. \] 

\mypar{Induction on the number of layers.} Ignoring the output activation, we first prove that
\begin{equation}\label{eq:induction_qgan}
\E_\phi \left[ \frac{1}{4^n} \sumxy D_\phi^{L}(x) \cdot D_\phi^{L}(y) \right] \geq \frac{1}{4} \prod_{l=1}^{L} \frac{m_l \sig_l^2 (1+\gamma)^2}{4} \,,
\end{equation}
by induction on $L$. Let $k$ be the unique index such that $\al_k = 1$. For the base case, we have
\begin{align*}
\E_\phi \left[ \frac{1}{4^n} \sumxy D_\phi^0(x) \cdot D_\phi^0(y) \right]
&= \frac{1}{4^n} \sumxy x \cdot y \\
&= \frac{1}{4^n}  \sum_{x', y'} (x' \cdot y'+1) - (x' \cdot y') - (x' \cdot y') + (x' \cdot y') \\
&= \frac{1}{4^n}  \sum_{x', y'} 1 = \frac{1}{4}
\end{align*}
where $x', y' \in \B^{n-1}$ are defined as $x, y \in \B^n$ with the $k$th component removed. Now for any fixed $l : 1 \to L$, we have
\begin{align*}
& \E_\phi \left[ \sumxy D_\phi^{l}(x) \cdot D_\phi^{l}(y) \right] \\
& \qquad \qquad = \sum_{i = 1}^{m_l} \E_\phi \left[ \sumxy R \left( A^l D_\phi^{l-1}(x) + B^l \right)_i R \left( A^l D_\phi^{l-1}(x) +B^l \right)_i \right] \,.
\end{align*}
To reduce this further, let us fix $i$ and simplify notation by defining
\begin{align*}
f(x) \coloneqq \left( A^l D_\phi^{l-1}(x) + B^l \right)_i = \sum_j A^l_{ij} D_\phi^{l-1}(x)_j + B^l_i  \,.
\end{align*}
Expanding the leaky-ReLU as
\[ R(x) = \max(x, \gamma x) = x(H(x) + \gamma H(-x)) \,, \]
where $H(x)$ is the Heaviside step function, the goal is now to simplify the expression
\begin{align*}
& \E_\phi \left[ \sumxy R(f(x)) R(f(y)) \right] \\
& \qquad \qquad = \E_\phi \left[ \sumxy f(x)f(y) \Big[ H(f(x)) + \gamma H(-f(x)) \Big] \Big[ H(f(y)) + \gamma H(-f(y)) \Big] \right] \,.
\end{align*}
Let us first make note of the trivial fact that
\[ x H(x) = \frac{1}{2}(x + \abs{x}) \,, \]
and extend this decomposition in two dimensions to obtain the less trivial identities
\begin{align*}
xy \Big[ H(x)H(y) + H(-x)H(-y) \Big] &= \frac{1}{2} \Big( xy + \abs{xy} \Big) \,, \\
xy \Big[ H(x)H(-y) + H(-x)H(y) \Big] &= \frac{1}{2} \Big( xy - \abs{xy} \Big) \,.
\end{align*}
Unfortunately, these equations alone will produce cross-terms whose expectations do not have closed-form solutions. Thankfully, we can sidestep this entirely by summoning the symmetry of parameter distributions, which imply
\[ \E_\phi\left[ f(x)f(y) H(f(x)) H(f(y)) \right] = \E_\phi\left[ f(x)f(y) H(-f(x)) H(-f(y)) \right]\,. \]
Combining these two sets of equations, we obtain
\begin{align*}
&\E_\phi \left[ \sumxy f(x)f(y) \Big[ H(f(x)) + \gamma H(-f(x)) \Big] \Big[ H(f(y)) + \gamma H(-f(y)) \Big] \right] \\
= \frac{1+\gamma^2}{2} &\E_\phi \left[ \sumxy f(x) f(y)  \Big[ H(f(x))H(f(y)) + H(-f(x))H(-f(y)) \Big] \right]  \\ \qquad + \gamma &\E_\phi \left[ \sumxy f(x) f(y) \Big[ H(f(x))H(-f(y)) + H(-f(x))H(f(y)) \Big] \right]  \\
= \frac{1+\gamma^2}{4} &\E_\phi \left[ \sumxy \Big[ f(x)f(y) + \big| f(x)f(y) \big| \Big] \right]  \\ \qquad + \frac{\gamma}{2} &\E_\phi \left[ \sumxy \Big[ f(x)f(y) - \big| f(x)f(y) \big| \Big] \right] \\
= \frac{(1+\gamma)^2}{4} &\E_\phi \left[ \sumxy f(x)f(y) \right] + \frac{(1-\gamma)^2}{4} \E_\phi \left[ \sumxy \big| f(x)f(y) \big| \right] \\
= \frac{(1+\gamma)^2}{4} &\E_\phi \left[ \sumxy f(x)f(y) \right] + \frac{(1-\gamma)^2}{4} \E_\phi \left[ \left(\sumx \big| f(x) \big| \right)^2 \right] \\
\geq \frac{(1+\gamma)^2}{4} &\E_\phi \left[ \sumxy f(x)f(y) \right] \,.
\end{align*}
We now use the fact that parameters are independent (and symmetrically distributed), along with the fact that any term where $x$ or $y$ does not appear must vanish in the signed sum, to obtain 
\begin{align*}
\E_\phi \left[ \sumxy f(x)f(y) \right] &= \E_\phi \left[ \sumxy \left( A^l D_\phi^{l-1}(x) + B^l \right)_i \left( A^l D_\phi^{l-1}(y) + B^l \right)_i \right] \\
& = \sum_{j, k}  \E_\phi \left[ \sumxy \Big( A^l_{ij} D_\phi^{l-1}(x)_j + B^l_i \Big) \Big( A^l_{ik} D_\phi^{l-1}(x)_k + B^l_k \Big) \right] \\
& = \sum_{j, k} \E_\phi \left[ A^l_{ij} A^l_{ik} \right] \E_\phi \left[ \sumxy D_\phi^{l-1}(x)_j D_\phi^{l-1}(x)_k \right] \\
& = \sum_j \sig_l^2 \E_\phi \left[ \sumxy D_\phi^{l-1}(x)_j D_\phi^{l-1}(x)_j \right] \\
& = \sig_l^2 \E_\phi \left[ \sumxy D_\phi^{l-1}(x) \cdot D_\phi^{l-1}(x) \right] \,.
\end{align*}
This holds for each $i$, so we obtain
\begin{align*}
& \E_\phi \left[ \sumxy D_\phi^{l}(x) \cdot D_\phi^{l}(y) \right] \\
& \geq \frac{\sig_l^2(1+\gamma)^2}{4} \sum_{i=1}^{m_l} \E_\phi \left[ \sumxy D_\phi^{l-1}(x) \cdot D_\phi^{l-1}(x) \right] \\
& = \frac{m_l \sig_l^2(1+\gamma)^2}{4}  \E_\phi \left[ \sumxy D_\phi^{l-1}(x) \cdot D_\phi^{l-1}(x) \right] \,.
\end{align*}
The induction is complete, so Equation \eqref{eq:induction_qgan} holds.\\

\mypar{(Output activation.)} It now remains to assimilate the output activation $F$. For both min-max and Wasserstein GANs, we can decompose $F$ into odd and even parts and obtain
\[ F(x) = F_\odd(x) + F_\even(x) = F'(0)x + F_\even(x) \,, \]
where $F'(0) = 1/2$ for min-max GANs and $F'(0) = 1$ for Wasserstein GANs. It follows that
\begin{align*}
& \E_\phi \left[ \sumxy D_\phi(x) D_\phi(y) \right] \\
= & \E_\phi \left[ \sumxy F \left( A^{L+1} D_\phi^L(x) + B^{L+1} \right) F \left( A^{L+1} D_\phi^L(y) + B^{L+1} \right) \right] \\
= F'(0)^2 & \E_\phi \left[ \sumxy  \left( A^{L+1} D_\phi^L(x) + B^{L+1} \right) \left( A^{L+1} D_\phi^L(y) + B^{L+1} \right) \right] \\
+ & \E_\phi \left[ \sumxy F_\even \left( A^{L+1} D_\phi^L(x) + B^{L+1} \right) F_\even \left( A^{L+1} D_\phi^L(y) + B^{L+1} \right) \right] \\
+ F'(0) & \E_\phi \left[ \sumxy \left( A^{L+1} D_\phi^L(x) + B^{L+1} \right) F_\even \left( A^{L+1} D_\phi^L(y) + B^{L+1} \right) \right] \\
+ F'(0) & \E_\phi \left[ \sumxy F_\even \left( A^{L+1} D_\phi^L(x) + B^{L+1} \right) \left( A^{L+1} D_\phi^L(y) + B^{L+1} \right) \right] \,.
\end{align*}
The second term is non-negative since it is the expectation of a squared quantity, while the third and fourth terms vanish by symmetry of the distributions on $A^{L+1}$ and $B^{L+1}$. More precisely, a change of variables $(A^{L+1}, B^{L+1}) \to (-A^{L+1}, -B^{L+1})$ is invariant under expectation, and gives
\begin{align*}
& \E_\phi \left[ \sumxy \left( -A^{L+1} D_\phi^L(x) - B^{L+1} \right) F_\even \left( -A^{L+1} D_\phi^L(y) - B^{L+1} \right) \right] \\
= - & \E_\phi \left[ \sumxy \left( A^{L+1} D_\phi^L(x) + B^{L+1} \right) F_\even \left( A^{L+1} D_\phi^L(y) + B^{L+1} \right) \right]
\end{align*}
since $F_\even$ is even. The third term is therefore equal to its opposite, and vanishes. Similarly for the fourth term. It remains only to expand the remaining first term:
\begin{align*}
\E_\phi \Big[ c_\al(\phi)^2 \Big] &= \E_\phi \left[ F'(0)^2 \sumxy D_\phi(x) D_\phi(y) \right] \\
& \geq F'(0)^2 \E_\phi \left[ \frac{1}{4^n} \sumxy  \left( A^{L+1} D_\phi^L(x) + B^{L+1} \right) \left( A^{L+1} D_\phi^L(y) + B^{L+1} \right) \right] \\
& = F'(0)^2 \sum_{j, k} \E_\phi \left[ A^{L+1}_j A^{L+1}_k \right] \E_\phi \left[ \frac{1}{4^n} \sumxy D_\phi^L(x)_j D_\phi^L(y)_k \right] \\
& = F'(0)^2 \sum_{j} \sig_{L+1}^2 \E_\phi \left[ \frac{1}{4^n} \sumxy D_\phi^L(x)_j D_\phi^L(y)_j \right] \\
& = F'(0)^2 \sig_{L+1}^2 \E_\phi \left[ \frac{1}{4^n} \sumxy D_\phi^L(x) \cdot D_\phi^L(y) \right] \,.
\end{align*}
Using Equation \eqref{eq:induction_qgan}, and remembering that $F'(0) \geq 1/2$, the proof is drawn to a close:
\[ \E_\phi \Big[ c_\al(\phi)^2 \Big] \geq \frac{\sig_{L+1}^2}{16} \prod_{l=1}^{L} \frac{m_l \sig_l^2 (1+\gamma)^2}{4} \,. \qedhere \]
\end{proof}

\newpage
\section{Robustness of Experiments}\label{app:averaged_experiments}

In this section, we check the statistical significance and robustness of our numerical experiments from Section \ref{sec:qgans-experiments}. 
On one hand, we verify that the results are representative of typical training behavior, i.e., consistent over various random seeds that determine initial parameters and sampled batches from the true and generated distributions. On the other, we check whether our results are robust under inexact gradients. For this purpose, we run the same training setup as in Section \ref{sec:qgans-experiments}, but average the results over 10 random seeds, and estimate the generator gradients using Simultaneous Perturbation Stochastic Approximation (SPSA) \cite{spall1992spsa}. This algorithm provides an unbiased estimator for the full gradient, which we average over 10 perturbation vectors, and is typically used in real hardware experiments thanks to its constant cost with respect to the number of parameters.\\

\begin{figure*}[h]
\centering
\begin{subfigure}[t]{0.49\linewidth}
\subcaption{Relative Entropy}
\includegraphics[width=\linewidth]{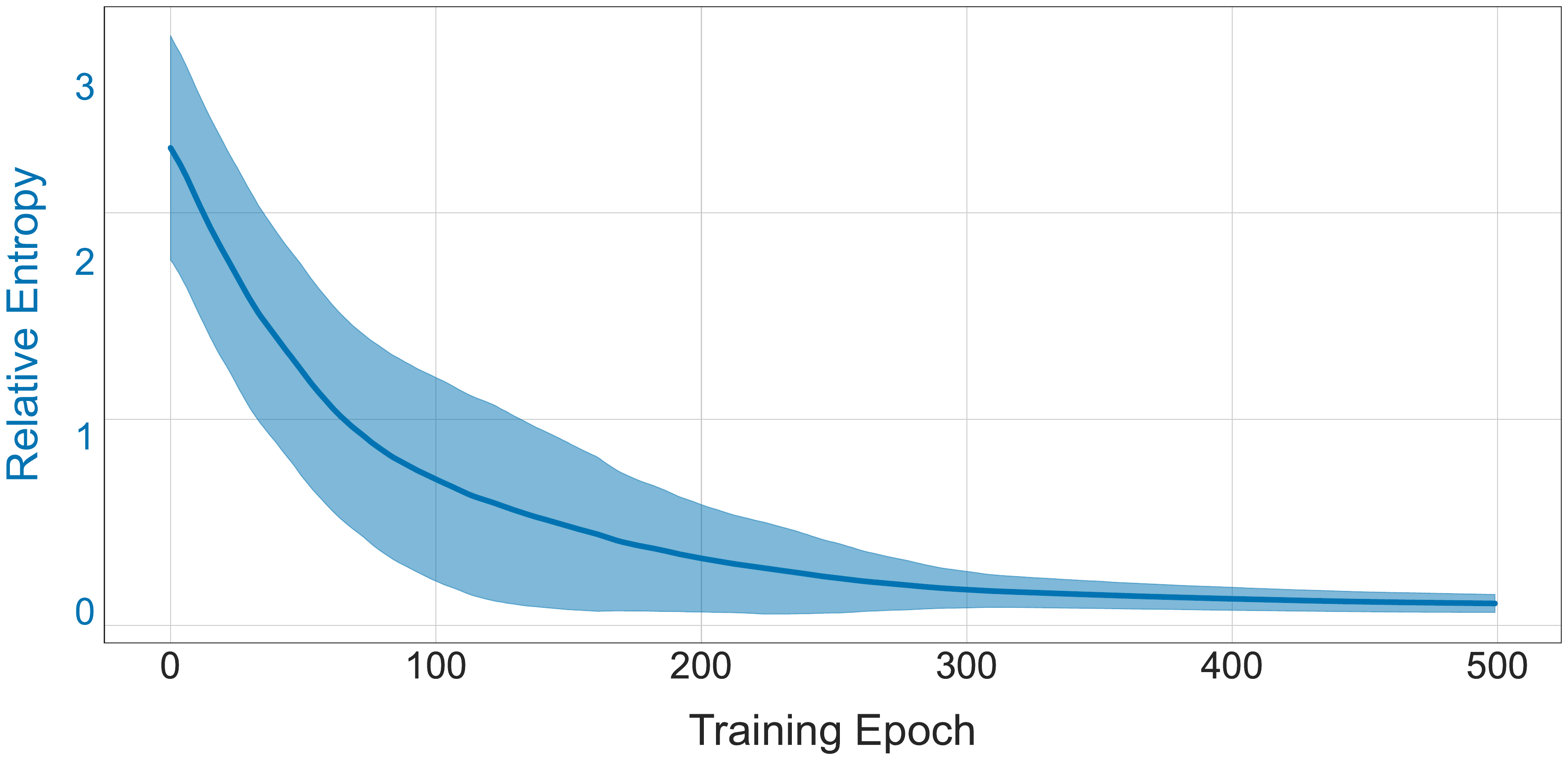} 
\end{subfigure} \hfill
\begin{subfigure}[t]{0.48\linewidth}
\subcaption{True \& Generated PDFs}\vspace{10pt}
\includegraphics[width=\linewidth]{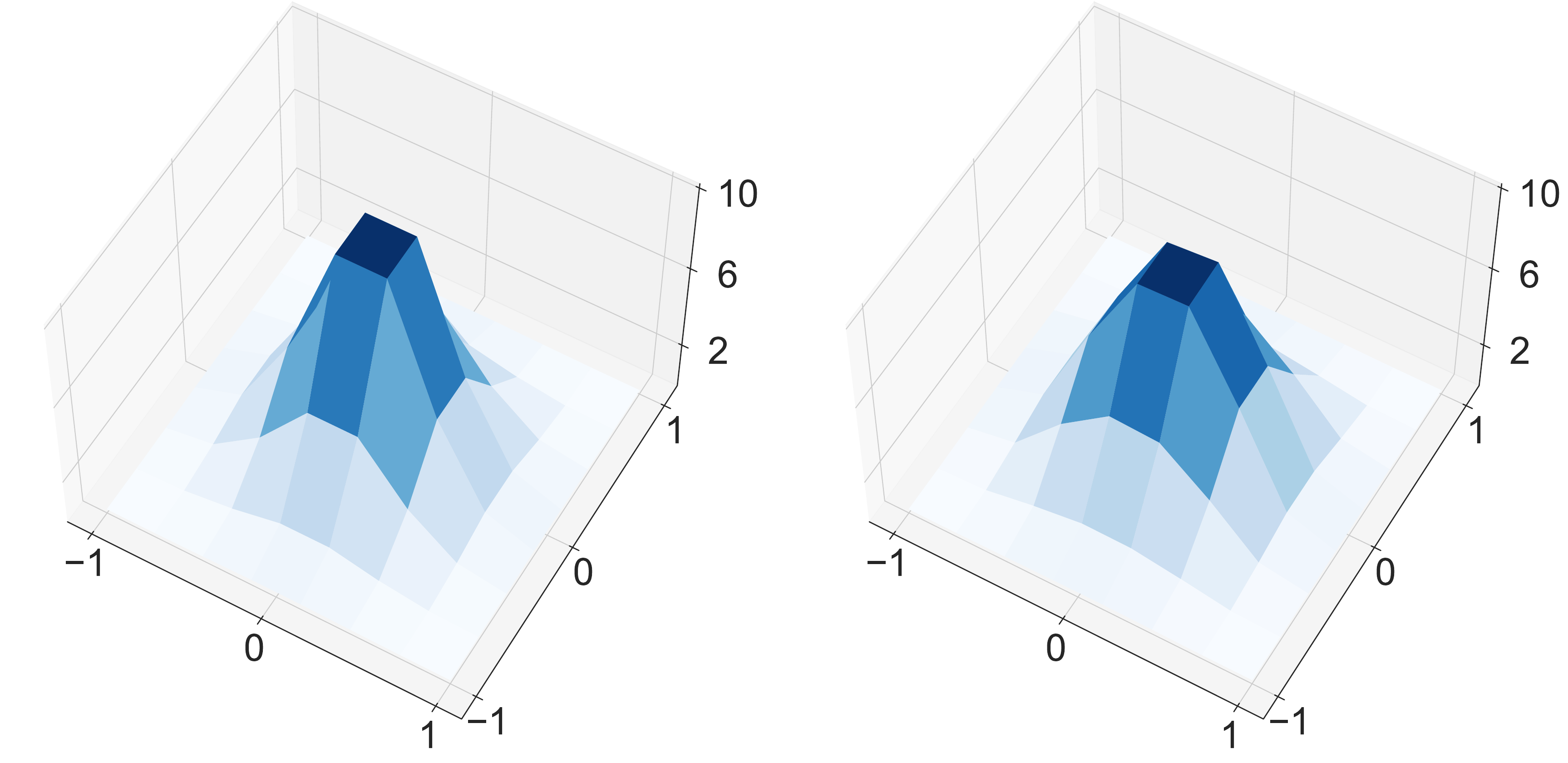}
\end{subfigure} \hfill
\begin{subfigure}[c]{0.49\linewidth}
\includegraphics[width=\linewidth]{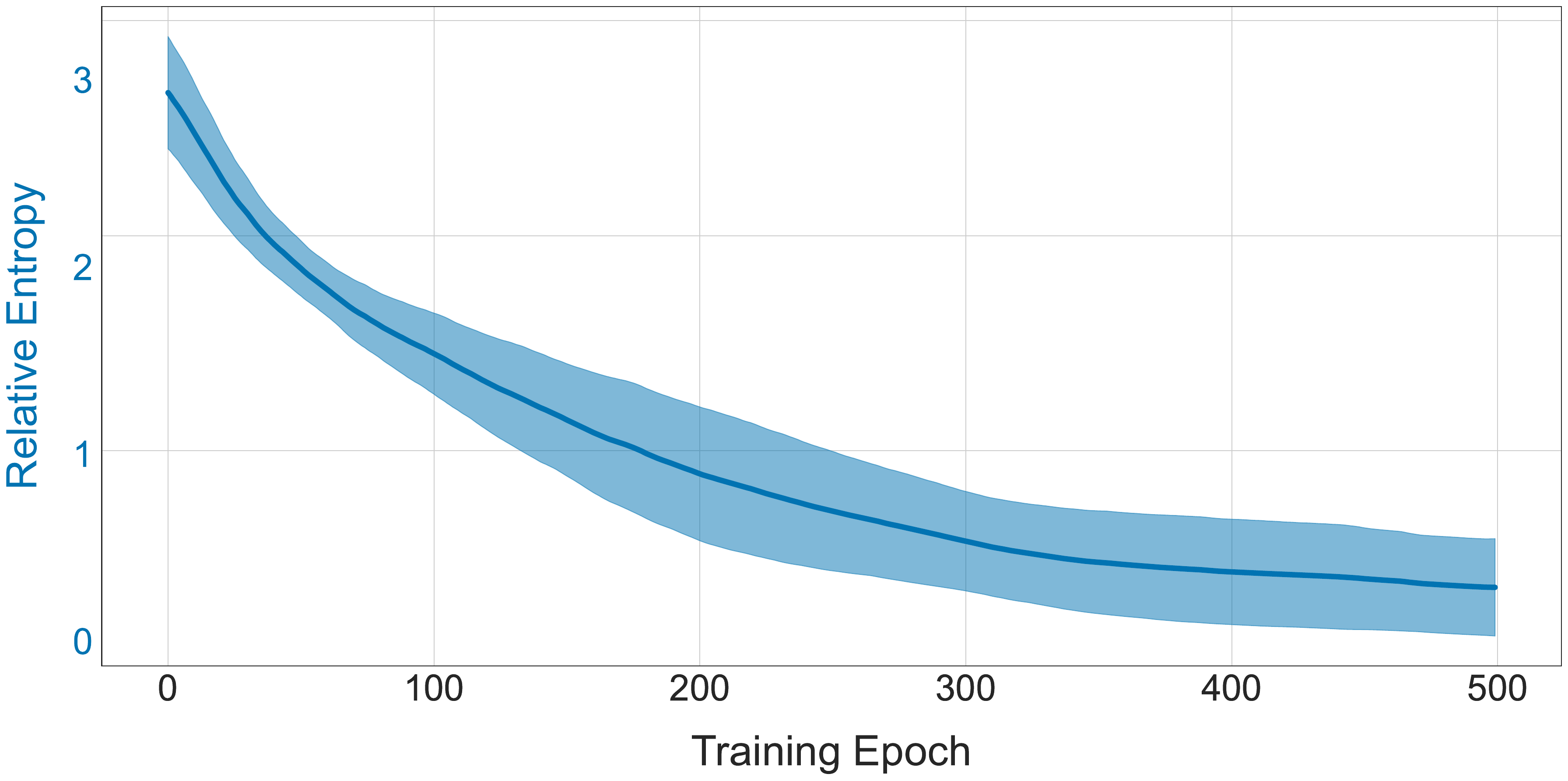} 
\end{subfigure} \hfill
\begin{subfigure}[c]{0.48\linewidth}
\includegraphics[width=\linewidth]{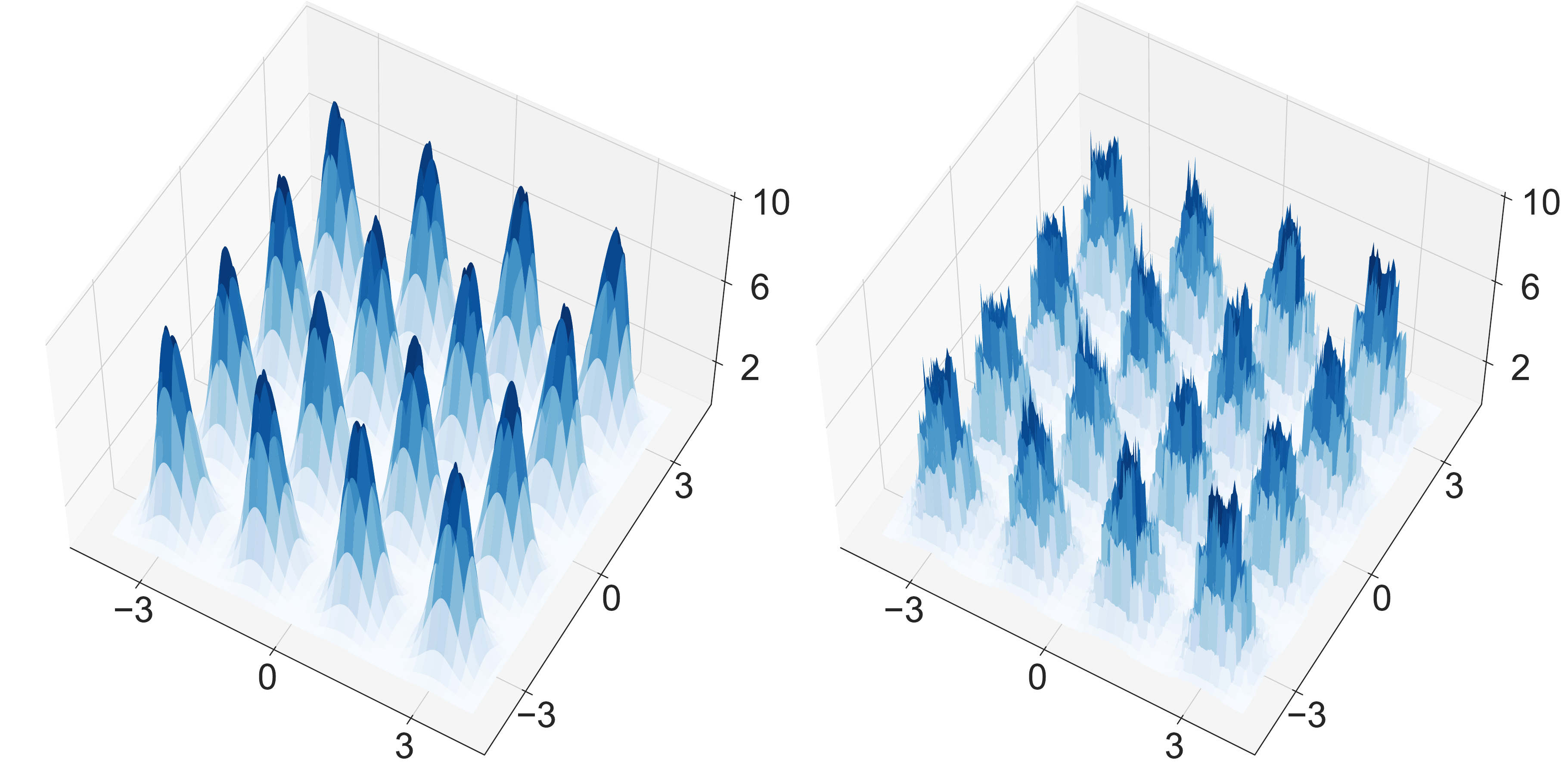} 
\end{subfigure} \hfill
\caption{Results for 6-qubit (top) and 16-qubit (bottom) averaged experiments, with gradients estimated using SPSA. (a): Relative entropy between true and generated distributions over the course of training. The entropy values are evaluated for 10 different random seeds. The solid lines represent the average over these seeds and the shaded regions represent the standard deviation. (b): True (left) and generated (right) probability density functions at the end of training -- once more averaged over 10 random seeds.}
\label{fig:results_stat_tests}
\end{figure*}

The results presented in Figure \ref{fig:results_stat_tests}(a) show the average evolution of the relative entropy over 500 training iterations, along with shaded standard deviations.
Overall, results are relatively robust, though the relative entropy converges more convincingly for 6 qubits than for 16 qubits. This behavior is reflected in the average entropy converging towards smaller values, but also in the decreasing standard deviation.
For 16 qubits, while the average entropy also decreases uniformly throughout training, the standard deviation stays more or less constant at later stages of training.
As discussed in the main text, this suggests that the model struggles to encode the underlying continuous distribution, possibly due to the lack of an inductive bias.
Finally, Figure \ref{fig:results_stat_tests}(b) illustrates the average of the learned PDFs for the various random seeds. While the 
6-qubit PDF is rather close to the target, the 16-qubit PDF is particularly spiky -- even in the averaged setting. This is in alignment with the relative entropy failing to converge to zero, and calls for further work to improve the capacity of quantum generators to encode continuous distributions effectively.

\vspace{4mm}

\end{document}